Exploratory Graph Analysis: A New Approach for Estimating the Number of Dimensions in Psychological Research.

Hudson F. Golino[1*], Sacha Epskamp[2]

[1]Graduate School of Psychology, Universidade Salgado de Oliveira, Rio de Janeiro, Brasil.

[2]University of Amsterdam, Netherlands.

* Corresponding author
E-mail: hfgolino@gmail.com

**Abstract**

The estimation of the correct number of dimensions is a long-standing problem in psychometrics. Several methods have been proposed, such as parallel analysis (PA), Kaiser-Guttman's eigenvalue-greater-than-one rule, multiple average partial procedure (MAP), the maximum-likelihood approaches that use fit indexes as BIC and EBIC and the less used and studied approach called *very simple structure* (VSS). In the present paper a new approach to estimate the number of dimensions will be introduced and compared via simulation to the traditional techniques pointed above. The approach proposed in the current paper is called *exploratory graph analysis* (EGA), since it is based on the *graphical lasso* with the regularization parameter specified using EBIC. The number of dimensions is verified using the *walktrap*, a random walk algorithm used to identify communities in networks. In total, 32,000 data sets were simulated to fit known factor structures, with the data sets varying across different criteria: number of factors (2 and 4), number of items (5 and 10), sample size (100, 500, 1000 and 5000) and correlation between factors (orthogonal, .20, .50 and .70), resulting in 64 different conditions. For each condition, 500 data sets were simulated using *lavaan*. The result shows that the EGA performs comparable to parallel analysis, EBIC, eBIC and to Kaiser-Guttman rule in a number of situations, especially when the number of factors was two. However, EGA was the only technique able to correctly estimate the number of dimensions in the four-factor structure when the correlation between factors were .7, showing an accuracy of 100% for a sample size of 5,000 observations. Finally, the EGA was used to estimate the number of factors in a real dataset, in order to compare its performance with the other six techniques tested in the simulation study.

**Introduction**

Estimating the number of dimensions in psychological and educational instruments is a long-standing problem in psychometrics [1, 2, 3]. Dimensions can be defined as the low set of features from a large set of correlated variables that collectively explain most of the variability in the original set [4], or as the underlying source of the variability presented in multivariate data [5]. Two main general traditions, within psychology, can be identified in the methods that have been proposed [2]. The first one examines patterns of eigenvalues, determining the number of factors based on some specified stopping rule. Two of the most known method following this tradition is the Kaiser-Guttman eigenvalue greater than one rule [6, 7] and Horn's parallel analysis [8]. The second general tradition compares the fit of structural models with varying numbers of factors and determines the number of factors to be retained based on the minimum average partial procedure (MAP) [9] or fit indexes such as the Bayesian information criterion (BIC) [10] and the extended Bayesian information criterion (EBIC) [11]. In addition to the above mentioned traditions, there is an alternative—less used and studied—approach called *very simple structure* (VSS) [12]. This approach assesses if the original correlation matrix can be reproduced by a simplified pattern matrix, in which the highest loading for each item is retained and the other loadings are set to zero.

In this paper we introduce a new approach to estimate the number of dimensions to be retained. We term this approach *exploratory graph analysis* (EGA), since it is based on estimating a graphical model [13, 14, 15] followed by cluster detection to estimate the number of dimensions in psychological data. EGA has the additional benefit over the abovementioned procedures that it not only estimates the *number* of dimensions but also which items belong to each dimension. We will compare this approach via simulation to the traditional or antique factor retention techniques: VSS, MAP, Kaiser-Guttman rule, parallel analysis, and the fit of different number of factors via BIC and EBIC. Finally, we have implemented EGA [16] in a free to use software package for the statistical programming language R.

Assessing Dimensionality in Psychological Data

Nowadays, psychology disposes of an impressive number of statistical procedures, with complex and flexible models carefully developed to deal with a multitude of problems. One may wonder whether estimating the number of dimensions using factor analysis still plays a role in the investigations, as it did some decades ago. The use of factor models is still very present as an early step in the process of construct validation [17], being considered "inexorably linked to the development of intelligence tests and to intelligence theory" (p. 37) [18]. A quick search in the Science Direct, an Elsevier web database for scientific publications, using the keywords "exploratory factor analysis" from 1990 to 2016 in journals from the fields of Arts and Humanities, Psychology and Social Sciences, yielded 40,132. From this total, 73.79% were published in the last ten years. So, as the very brief and non-systematic search pointed above shows, going in the same direction of previous papers [19], factor analysis is still widely used and broadly applied. However, reviews show that from 22% to 28% of papers published using exploratory factor analysis failed to report the specific extraction method used [32]. This is a very serious issue, because the extraction method used can impact the number of dimensions estimated. As will be pointed in the next paragraphs, each technique has its benefits and pitfalls, so reporting which method was used is extremely important.

Why does psychology need a new way to estimate the number of dimensions? The answer lies in the several studies published about the performance of the parallel analysis [18, 20, 21, 22, 23, 24, 25], the MAP [18, 24, 26], the *BIC* [27, 28, 29, 30] and the Kaiser-Guttman eigenvalue rule [2, 18, 24, 25, 31] in estimating the correct number of factors. In this line of research, it has been shown that parallel analysis and the MAP work quite well when there is a low or moderate correlation between factors, when the sample size is equal to or greater than 500 and when the factor loadings are from moderate to high [18, 21, 22]. However, they tend to underestimate the number of factors when the correlations between factors are high, when the sample size is small and when there is small number of indicators per factor [2, 18, 21, 22].

The Kaiser-Guttman rule is the default method for choosing the number of factors in many commercial software packages [32]. However, simulation studies show that this method overestimates the number of factors, especially with a large number of items and a large sample size [2, 18, 24, 25, 31]. Ruscio and Roche [2] provided startling evidence in this direction: the Kaiser-Gutman rule overestimated the number of factors in 89.87% of the 10,000 simulated datasets, generated with different number of factors, sample size, number of items, number of response categories per item and strength of correlation between factors. In face of the evidences from the simulation studies, some researchers strongly recommend not to use this method [24, 32].

Regarding the BIC, evidences are contrasting. Preacher, Zhang, Kim and Mels [29] showed that BIC performs well when the sample size is small, but tends to overestimate the number of factors in large datasets. However, Dziak, Coffman, Lanza and Li [27] showed that BIC decreases its underestimation and increases its correctness in estimating the number of factors when the sample size is greater than 200 cases.

It is important to point that, to our best knowledge, there is no study showing how the very simple structure approach behaves under different conditions. These simulation studies highlight a very complicated problem within psychology, since it is very common to find areas in which the correlation between factors is high, especially in the intelligence field [18]. Thus, in such situations, parallel analysis, MAP and comparing different number of factors via BIC perform bad, in average, proving that estimating the number of factors is still a non-trivial task, in spite of the past decades' developments. It seems that Kaiser's dictum remains valid: "a solution to the number-of-factors problem in factor analysis is easy... But the problem, of course is to find *the* solution" [33].

The next section will introduce a new approach to estimate the number of dimensions, called *exploratory graph analysis* (EGA). EGA will be compared to Parallel Analaysis, MAP, BIC, EBIC, Kaiser-Guttman rule and VSS in a simulation study with 32,000 simulated data sets, created by 64 conditions varying in four different criteria: number of factors (2 and 4), number of items per factor (5 and 10), sample size (100, 500, 1000 and 5000) and correlation between factors (orthogonal, .20, .50 and .70). In the last section, EGA will be used to estimate the number of factors from a real dataset with an empirically found factor structure. This will enable the comparison of EGA with the six techniques tested in the simulation study.

## Network Psychometrics

Recent literature has focused on the estimation of undirected network models, so called Markov Random Fields [13] to psychological datasets. In these network models, nodes represent random variables (as opposed to e.g., people in social networks) which are connected by edges or links indicating the level of interaction between these variables. These models focus on the estimation of direct relationships between observed variables rather than modeling observed variables as functions of latent common causes. Such models have shown great promise in diverse psychological fields such as psychopathology [34 35, 36, 37, 38], attitude formation [39] quality of life research [40] and developmental psychology [41]. Forming a network structure on psychological data, however, is not an easy task. The field of *network psychometrics* [42] emerged as a response to these concerns with the estimation of such network models.

The network model we will utilize in this paper is termed the Gaussian graphical model (GGM) [13] which models multivariate normally distributed network directly through the inverse covariance matrix. Each element of the inverse covariance matrix corresponds to a connection, an *edge,* in the network, linking two variables, *nodes*, if they feature a pairwise interaction. These edges can be standardized, visualized and more easily interpreted as *partial correlation coefficients* of two variables after conditioning on all other variables in the dataset. Partial correlation coefficients of exactly zero indicate that there is no edge between two nodes. Thus, in a GGM, if two variables are not connected, they are conditionally independent after conditioning on all other variables in the network (see FigFig 7, for an example).

While a GGM can be estimated directly by inverting the sample variance-covariance matrix, doing so can lead to large standard errors and unstable parameter estimates in relatively small datasets (i.e., typical sample sizes in psychological research) due to overfitting. A popular technique used in estimating GGMs is to not directly invert the variance-covariance matrix but to estimate this model using penalized maximum likelihood estimation. In particular, the *least absolute shrinkage and selection operator* (LASSO) [43] can be used to estimate a GGM while guarding against overfitting. Using the LASSO has the additional benefits in that it can result in many parameters to be estimated to exactly equal zero; indicating conditional independence and increasing interpretability of a network structure. Because of these properties, LASSO estimation has become the go-to estimation method for network models on psychological datasets [e.g. 38, 40, 44]. When using LASSO estimation one needs to set a tuning parameter that loosely controls the sparsity of the resulting network structure. A typical way of setting this tuning parameter is by estimating a model on 100 different tuning parameters and selecting the value that minimizes some criterion. For GGM estimation, minimizing the extended Bayesian information criterion [11] has been shown to work well in retrieving the true network structure [15]. This methodology has been implemented in the *qgraph* R package [45, 46] for easy usage on psychometric datasets.

**Exploratory Graph Analysis**. The modeling of psychological datasets through network models originates with the work of van der Maas et al. [41], who show that a dataset that corresponds to a general factor model can be simulated using a fully connected network model as well. A section of a network in which all nodes are fully connected is also termed a *clique*, and a section in which many nodes are connected with each other is termed a *cluster*. Such clusters are of particular interest to psychometrics, as it is argued clusters of nodes will lead to comparable data as a latent variable model, or, depending on one's assumptions on the underlying causal structure, influence due to latent variables will manifest in network structures as such clusters or even cliques in which all nodes interact with each other. For instance, in psychopathological literature it is argued that clusters of nodes representing symptoms correspond to psychopathological disorders [34, 35]. Similar arguments have been made for stable personality traits, which routinely come up as clusters in an estimated network structure [45, 47, 48].

The relationship between latent variables on the one hand and network clusters on the other goes deeper than mere philosophical speculation and empirical findings. It can directly be seen that if a latent variable model is the true underlying causal model, we would expect indicators in a network model to feature strongly connected clusters for each latent variable. Since edges correspond to partial correlation coefficients between two variables after conditioning on all other variables in the network, and two indicators cannot become independent after conditioning on observed variables given that they are both caused by a latent variable, the edge strength between two indicators should not be zero. In fact, network models can be shown to be mathematically equivalent under certain conditions to latent variable models in both binary [42, 49] and Gaussian datasets [50], in which case each latent variable is represented by a rank-1 cluster. Thus, when defining a cluster as a group of connected nodes regardless of edge weight, we can state the following relationship as a fundamental rule of network psychometrics: Clusters in network = latent variables.

It should be noted that when multiple correlated latent variables underlie distinct sets of indicators, none of the edges should be missing as we technically cannot condition on any observed variable to make two indicators independent. However, we would expect the partial correlation between two indicators of the same latent variable to be much stronger than the partial correlation between two indicators of different latent variables. Furthermore, when using LASSO estimation, we would expect these already small edge weights to be pushed more easily to zero simply due to the penalization. As such, we expect an algorithm to detect weighted network clusters to indicate indicators of the same latent variable.

In a more mathematical point of view, let $\boldsymbol{y}$ represent a centered random vector of $K$ responses, which we assume to be multivariate normally distributed with some variance-covariance matrix $\boldsymbol{\Sigma}$:

$$\boldsymbol{y} \sim N_P\left(\boldsymbol{0}, \boldsymbol{\Sigma}\right).$$

In factor analysis, we typically assume that the response of subject $p$ is caused through a linear factor model by a set of $M$ latent variables $\boldsymbol{\eta}$ plus random error $\varepsilon$:

$$\boldsymbol{y_p} = \boldsymbol{\Lambda}\boldsymbol{\eta_P} + \varepsilon_P,$$

in which $\boldsymbol{\Lambda}$ is a $K \times M$ factor loadings matrix. This leads to the well known factor analysis model:

$$\boldsymbol{\Sigma} = \boldsymbol{\Lambda}\boldsymbol{\psi}\boldsymbol{\Lambda}^\top + \boldsymbol{\Theta},$$

in which $\boldsymbol{\Theta} = \text{Var}\left(\varepsilon\right)$ and $\boldsymbol{\psi} = \text{Var}\left(\boldsymbol{\eta}\right)$. Often it is assumed that each item only loads on one factor (simple structure), and thus that $\boldsymbol{\Lambda}$ can be reordered to be block diagonal:

$$\boldsymbol{\Lambda} = \begin{bmatrix} \lambda_1 & 0 & \cdots & 0 \\ 0 & \lambda_2 & \cdots & 0 \\ \vdots & \vdots & \ddots & \vdots \\ 0 & 0 & \cdots & \lambda_M \end{bmatrix},$$

Furthermore, we assume $\boldsymbol{\Theta}$ to be diagonal (local independence):

$$\boldsymbol{\Theta} = \begin{bmatrix} \Theta_1 & 0 & \cdots & 0 \\ 0 & \Theta_2 & \cdots & 0 \\ \vdots & \vdots & \ddots & \vdots \\ 0 & 0 & \cdots & \Theta_K \end{bmatrix}.$$

In network modeling, the Gaussian graphical model [13] is used in which the inverse variance-covariance matrix is modeled [51]:

$$\boldsymbol{\Sigma} = \boldsymbol{K}^{-1}.$$

A zero element in $\boldsymbol{K}$ indicates conditional independence:

$$y_i \perp\!\!\!\perp y_j \mid \boldsymbol{y}^{-(ij)} \iff k_{i,j} = k_{j,i} = 0,$$

in which $\boldsymbol{y}^{-(ij)}$ indicates $\boldsymbol{y}$ without elements $i$ and $j$, and negative elements of $\boldsymbol{K}$ can be standardized to equal *partial correlation coefficients*:

$$\rho_{ij} = -\frac{k_{ij}}{\sqrt{k_{ii}k_{jj}}}.$$

These partial correlation coefficients are thus proportional to the inverse variance-covariance matrix, and can be used to form a partial correlation network. As such, off-diagonal elements of $\boldsymbol{K}$ encode a network structure.

Relating the above expressions and applying the Woodbury matrix identity we obtain:

$$\boldsymbol{K} = \left(\boldsymbol{\Lambda}\boldsymbol{\psi}\boldsymbol{\Lambda}^\top + \boldsymbol{\Theta}\right)^{-1} = \boldsymbol{\Theta}^{-1} - \boldsymbol{\Theta}^{-1}\boldsymbol{\Lambda}(\boldsymbol{\psi}^{-1} + \boldsymbol{\Lambda}^\top\boldsymbol{\Theta}^{-1}\boldsymbol{\Lambda})^{-1}\boldsymbol{\Lambda}^\top\boldsymbol{\Theta}^{-1}.$$

Since $\boldsymbol{\Theta}$ is diagonal, so is $\boldsymbol{\Theta}^{-1}$, leading to $\boldsymbol{\Theta}^{-1}\boldsymbol{\Lambda}$ to be block diagonal and $\boldsymbol{\Lambda}^\top\boldsymbol{\Theta}^{-1}\boldsymbol{\Lambda}$ to be diagonal. Let $\boldsymbol{X} = (\boldsymbol{\psi}^{-1} + \boldsymbol{\Lambda}^\top\boldsymbol{\Theta}^{-1}\boldsymbol{\Lambda})^{-1}$. Then, $\boldsymbol{K}$ becomes a block matrix in which every block is constructed of the inner product of factor loadings and inverse residual variances, every diagonal block is scaled by diagonal elements of $\boldsymbol{X}$ and every off-diagonal block is scaled by off-diagonal values of $\boldsymbol{X}$.

Since $\boldsymbol{\psi}$ must be positive definite it follows that $\boldsymbol{X}$ must be positive definite as well. Typically in factor analysis the first factor loadings or the latent variance-covariances are fixed to 1 to identify the model. We can, however, without loss of information, also constrain the diagonal of $\boldsymbol{X}$ to equal 1. It then follows that every absolute off-diagonal value of $\boldsymbol{X}$ must be smaller than 1. From the formation of $\boldsymbol{X}$ follows that off-diagonal values of $X$ equal zero if the latent factors are orthogonal. Hence, the above decomposition shows that:

1. If the latent factors are orthogonal, the resulting GGM consists of unconnected clusters.

2. Assuming factor loadings and residual variances are reasonably on the same scale for every item, the off-diagonal blocks of $K$ will be scaled closer to zero than the diagonal blocks of $K$. Hence, the resulting GGM will contain weighted clusters for each factor.

This line of reasoning leads us to develop Exploratory Graph Analysis (EGA), in which firstly we estimate the correlation matrix of the observable variables, then the graphical LASSO estimation is used to obtain the sparse inverse covariance matrix, with the regularization parameter defined via EBIC over 100 different values. Finally, the *walktrap* algorithm [52] is used to find the number of dense subgraphs (communities or clusters) of the partial correlation matrix computed in the previous step. The *walktrap* algorithm provides a measure of similarities between vertices based on random walks which can capture the community/cluster structure in a graph [52]. The number of clusters identified equals the number of latent factors in a given dataset.

In sum, we expect EGA to present a high accuracy in estimating the number of dimensions in psychology-like datasets due to the use of the LASSO technique [43]. Partial correlation is one of the methods used to estimate network models, but it suffers from an important issue: even when two variables are conditionally independent, the estimated partial correlation coefficient is not zero due to sampling variation [46]. In other words, partial correlation can reflect spurious correlations. This issue can be solved using regularization techniques, such as the LASSO [43], which is one of the most prominent methods for network estimation on psychological datasets [38, 40, 44]. When LASSO is used to estimate a network, it avoids overfitting by shrinking the partial correlation coefficients, so small coefficients are estimated to be exactly zero, indicating conditional independence and making the interpretability of the network structure easier [46]. Since the LASSO can be used to control spurious connections, it is reasonable to expect it will provide high accurate estimates of the underlying structure of the data when combined with a community detection algorithm such as the walktrap [52].

**Simulation Study**

***Method***

*Design*

Three thousand two hundred data sets were simulated to fit known factor structures, with the data sets varying across different criteria. The data generation design manipulated four variables, number of factors (2 and 4), number of items per factor (5 and 10), sample size (100, 500, 1000 and 5000) and correlation between factors (orthogonal, .20, .50 and .70), in a total of 64 different conditions with a 2x2x4x4 design. For each condition, 500 data sets were simulated using the R [53] package *lavaan* [54], resulting in the above mentioned 32,000 data sets. The simulated data came from a centered multivariate normal distribution, with factor loadings and variances set to unity, and every item artificially dichotomized at their respective theoretical mean zero.

Each factor was composed by five or ten dichotomous items. The choice of using this kind of items can be justified by the dichotomous nature of a significant number of intelligence test items, especially those requiring the respondents to perform some task with only one correct answer, such as the Raven's progressive matrices [55], the Wiener Matrizen-Test 2 [56] or the more recent tests from the International Cognitive Ability Resource [57, 58]. Since high correlation between factors are often found in intelligence researches, we have intended to mimic the nature of the field, so the comparison between the proposed exploratory graphical analysis and the traditional/antique techniques are easier to understand and to interpret.

*Data Analysis*

The simulated data sets were submitted to seven different methods to estimate the number of dimensions (factors): (1) very simple structure (VSS) [12] with complexity 1; (2) minimum average partial procedure (MAP) [9]; (3) the fit of different number of factors, from 1 to 10, via BIC; (4) the fit of different number of factors, from 1 to 10, via EBIC; (5) Horn's Parallel Analysis (PA) [8] using the generalized weighted least squares factor method; (6) Kaiser-Guttman eigenvalue greater than one rule [6, 7]; and (7) Exploratory Graph Analysis. The first five methods were implemented using the *R* package *psych* [59]. Since the items are dichotomous, the PA was applied using tetrachoric correlations for the real and simulated data. The eigenvalue greater than one rule was applied taking the observed eigenvalues calculated during the PA procedure.

The exploratory graph analysis was applied using the R package *EGA* [16]. This package has a function named *EGA* with two arguments: *data* and *plot.EGA*. The first one is used to specify the dataset and the second one is a logical argument, if *TRUE* returns a network showing the dimensions estimated. The *EGA* function returns a list with 5 elements: *ndim* (number of dimensions estimated), *correlation* (a matrix of zero-order correlation between the items), *glasso* (a matrix with the partial correlation estimated using EBICglasso, from *qgraph*), *wc* (the *walktrap community* membership of the items), *dim.variables* (a dataframe with two columns: items and their respective estimated dimension). The *EGA* function firstly calculates the polychoric correlations via the *cor_auto* function of the *qgraph* package [45]. Secondly, the function uses the *EBICglasso* from the *qgraph* package [45] to estimate the sparse inverse covariance matrix with the graphical lasso technique. The *EBICglasso* function runs one hundred values of the regularization parameter, generating one hundred graphs. The EBIC is computed and the graph with the smallest EBIC is selected. Finally, the *EGA* function uses the *walktrap* algorithm [52] to find the number of dense subgraphs (communities) of the partial correlation matrix computed in the previous step, via the *walktrap.community* function available in *igraph* [60]. The walktrap algorithm provides a measure of similarities between vertices based on random walks which can capture the community structure in a graph [52].

Three indexes were recorded for each one of the 32,000 datasets, following Garrido, Abad and Posada [17]. The first index is the accuracy to correctly recover the number of factors. For example, in the four factor structure the accuracy equals one if four factors are estimated and zero otherwise. So it is possible to compute descriptive statistics based on the accuracy of each method in each group of 500 simulated data sets, for each condition. The second index, bias error, is the difference between the number of factors estimated and the true number of factors. A positive bias error indicates that the method is overestimating the number of factors. On the other hand, a negative bias error indicates that the method is underestimating the number of factors, and a bias error of zero indicates a complete lack of bias. The mean bias error (MBE) is calculated as the sum of the bias error divided by the number of datasets generated for each condition. The third index is the absolute error, which is the absolute value of the bias error. The mean absolute error (MAE) is calculated as the sum of the absolute error divided by the number of datasets generated for each condition. As pointed by Garrido, Abad and Posada [17], the bias error cannot be used alone for verifying the precision of a method to estimate the number of factors, since errors of under- and overfactoring can compensate each other. This does not happen with the accuracy index or with the absolute error index. A mean absolute error of zero indicates a perfect accuracy, while higher values are evidence of deviation from the correct number of dimensions.

## **Results**

*Structure with Two Factors*

      *Accuracy*:

      Table 1 shows the mean accuracy and its standard deviation for each method, in each condition. When the correlation between factors was zero (orthogonal) the methods presented a mean accuracy ranging from 98% to 100%, except for the VSS method, which presented a mean accuracy of 31% (SD = 46%). As the sample size increased, the mean accuracy of VSS decreased from 76% (sample size of 100) to 3% (sample size of 5,000). On the other side, all the other methods achieved a mean accuracy of 100% for sample sizes of 500, 1000 and 5000. The exactly same pattern appeared when the correlation was .2. When the correlation between factors was .5, BIC, eBIC, Kaiser-Guttman's eigenvalue rule, PA and EGA presented an overall mean accuracy greater than 90%, while VSS presented a mean accuracy of 22% and the MAP 67%. When the correlation was high (.7), only eBIC, PA and EGA presented an overall mean accuracy greater than 90%. In general, the increase in the number of items per factor lead to an increase in the mean accuracy and a decrease in the standard deviation, especially in the high correlation scenario (Table 1).

**Table 1**
Mean accuracy and its standard deviation, for each method and each condition, for the two factor structure. VSS = Very Simple Structure;
BIC = Bayesian Information Criteria; EBIC = Extended Bayesian Information Criteria; MAP = Minimum Average Partial
procedure; Kaiser = Kaiser-Guttman eigenvalue greater than one rule; PA = Parallel Analysis; EGA = Exploratory Graph Analysis.
Low correlation = .2; Moderate Correlation = .5; High Correlation = .7. The rows show the aggregate mean and standard deviation for each level of
correlation (bold), sample size (bold and italicized) and number of items per factor (non-italicized).

| | VSS | | MAP | | BIC | | EBIC | | Kaiser | | PA | | EGA | | Total | |
|---|---|---|---|---|---|---|---|---|---|---|---|---|---|---|---|---|
| | Mean | SD | Mean | SD | Mean | SD | Mean | SD | Mean | SD | Mean | SD | Mean | SD | Mean | SD |
| **Orthogonal** | **0.31** | **0.46** | **0.99** | **0.10** | **0.99** | **0.07** | **1.00** | **0.02** | **1.00** | **0.03** | **0.99** | **0.08** | **0.98** | **0.13** | **0.90** | **0.31** |
| *100* | *0.76* | *0.43* | *0.96* | *0.21* | *0.98* | *0.14* | *1.00* | *0.03* | *1.00* | *0.06* | *0.97* | *0.17* | *0.94* | *0.25* | *0.94* | *0.23* |
| 5 | 0.71 | 0.45 | 0.91 | 0.28 | 0.96 | 0.20 | 1.00 | 0.04 | 1.00 | 0.04 | 0.95 | 0.22 | 0.92 | 0.27 | 0.92 | 0.27 |
| 10 | 0.81 | 0.39 | 1.00 | 0.00 | 1.00 | 0.00 | 1.00 | 0.00 | 0.99 | 0.08 | 1.00 | 0.06 | 0.95 | 0.22 | 0.96 | 0.19 |
| *500* | *0.36* | *0.48* | *1.00* | *0.00* | *1.00* | *0.00* | *1.00* | *0.00* | *1.00* | *0.00* | *1.00* | *0.03* | *1.00* | *0.00* | *0.91* | *0.29* |
| 5 | 0.54 | 0.50 | 1.00 | 0.00 | 1.00 | 0.00 | 1.00 | 0.00 | 1.00 | 0.00 | 1.00 | 0.04 | 1.00 | 0.00 | 0.93 | 0.25 |
| 10 | 0.18 | 0.38 | 1.00 | 0.00 | 1.00 | 0.00 | 1.00 | 0.00 | 1.00 | 0.00 | 1.00 | 0.00 | 1.00 | 0.00 | 0.88 | 0.32 |
| *1000* | *0.08* | *0.27* | *1.00* | *0.00* | *1.00* | *0.00* | *1.00* | *0.00* | *1.00* | *0.00* | *1.00* | *0.00* | *1.00* | *0.00* | *0.87* | *0.34* |
| 5 | 0.08 | 0.28 | 1.00 | 0.00 | 1.00 | 0.00 | 1.00 | 0.00 | 1.00 | 0.00 | 1.00 | 0.00 | 1.00 | 0.00 | 0.87 | 0.34 |
| 10 | 0.07 | 0.26 | 1.00 | 0.00 | 1.00 | 0.00 | 1.00 | 0.00 | 1.00 | 0.00 | 1.00 | 0.00 | 1.00 | 0.00 | 0.87 | 0.34 |
| *5000* | *0.03* | *0.18* | *1.00* | *0.00* | *1.00* | *0.00* | *1.00* | *0.00* | *1.00* | *0.00* | *1.00* | *0.00* | *1.00* | *0.00* | *0.86* | *0.34* |
| 5 | 0.00 | 0.00 | 1.00 | 0.00 | 1.00 | 0.00 | 1.00 | 0.00 | 1.00 | 0.00 | 1.00 | 0.00 | 1.00 | 0.00 | 0.86 | 0.35 |
| 10 | 0.07 | 0.25 | 1.00 | 0.00 | 1.00 | 0.00 | 1.00 | 0.00 | 1.00 | 0.00 | 1.00 | 0.00 | 1.00 | 0.00 | 0.87 | 0.34 |
| **Low Correlation** | **0.33** | **0.47** | **0.98** | **0.12** | **0.99** | **0.10** | **1.00** | **0.03** | **1.00** | **0.04** | **0.99** | **0.10** | **0.99** | **0.10** | **0.90** | **0.30** |
| *100* | *0.79* | *0.40* | *0.94* | *0.24* | *0.96* | *0.19* | *1.00* | *0.05* | *0.99* | *0.08* | *0.96* | *0.20* | *0.96* | *0.19* | *0.94* | *0.23* |
| 5 | 0.71 | 0.45 | 0.88 | 0.33 | 0.93 | 0.26 | 0.99 | 0.08 | 0.99 | 0.09 | 0.93 | 0.26 | 0.96 | 0.20 | 0.91 | 0.28 |
| 10 | 0.88 | 0.33 | 1.00 | 0.04 | 1.00 | 0.04 | 1.00 | 0.06 | 0.99 | 0.08 | 0.99 | 0.08 | 0.96 | 0.19 | 0.98 | 0.16 |
| *500* | *0.36* | *0.48* | *1.00* | *0.03* | *1.00* | *0.00* | *1.00* | *0.00* | *1.00* | *0.00* | *1.00* | *0.00* | *1.00* | *0.00* | *0.91* | *0.29* |
| 5 | 0.54 | 0.50 | 1.00 | 0.04 | 1.00 | 0.00 | 1.00 | 0.00 | 1.00 | 0.00 | 1.00 | 0.00 | 1.00 | 0.00 | 0.93 | 0.25 |
| 10 | 0.18 | 0.39 | 1.00 | 0.00 | 1.00 | 0.00 | 1.00 | 0.00 | 1.00 | 0.00 | 1.00 | 0.00 | 1.00 | 0.00 | 0.88 | 0.32 |
| *1000* | *0.12* | *0.33* | *1.00* | *0.00* | *1.00* | *0.00* | *1.00* | *0.00* | *1.00* | *0.00* | *1.00* | *0.00* | *1.00* | *0.00* | *0.87* | *0.33* |
| 5 | 0.14 | 0.35 | 1.00 | 0.00 | 1.00 | 0.00 | 1.00 | 0.00 | 1.00 | 0.00 | 1.00 | 0.00 | 1.00 | 0.00 | 0.88 | 0.33 |
| 10 | 0.10 | 0.30 | 1.00 | 0.00 | 1.00 | 0.00 | 1.00 | 0.00 | 1.00 | 0.00 | 1.00 | 0.00 | 1.00 | 0.00 | 0.87 | 0.34 |
| *5000* | *0.03* | *0.16* | *1.00* | *0.00* | *1.00* | *0.00* | *1.00* | *0.00* | *1.00* | *0.00* | *1.00* | *0.00* | *1.00* | *0.00* | *0.86* | *0.35* |
| 5 | 0.00 | 0.00 | 1.00 | 0.00 | 1.00 | 0.00 | 1.00 | 0.00 | 1.00 | 0.00 | 1.00 | 0.00 | 1.00 | 0.00 | 0.86 | 0.35 |
| 10 | 0.05 | 0.23 | 1.00 | 0.00 | 1.00 | 0.00 | 1.00 | 0.00 | 1.00 | 0.00 | 1.00 | 0.00 | 1.00 | 0.00 | 0.86 | 0.34 |
| **Moderate Correlation** | **0.22** | **0.42** | **0.67** | **0.47** | **0.92** | **0.27** | **0.98** | **0.13** | **0.93** | **0.26** | **0.98** | **0.15** | **0.96** | **0.19** | **0.81** | **0.39** |
| *100* | *0.39* | *0.49* | *0.68* | *0.47* | *0.67* | *0.47* | *0.93* | *0.26* | *0.86* | *0.34* | *0.91* | *0.29* | *0.86* | *0.35* | *0.76* | *0.43* |
| 5 | 0.39 | 0.49 | 0.37 | 0.48 | 0.49 | 0.50 | 0.87 | 0.34 | 0.75 | 0.43 | 0.83 | 0.37 | 0.85 | 0.35 | 0.65 | 0.48 |
| 10 | 0.39 | 0.49 | 0.99 | 0.12 | 0.86 | 0.35 | 0.99 | 0.09 | 0.98 | 0.15 | 0.98 | 0.14 | 0.86 | 0.35 | 0.86 | 0.35 |
| *500* | *0.33* | *0.47* | *0.70* | *0.46* | *1.00* | *0.00* | *1.00* | *0.00* | *0.91* | *0.28* | *1.00* | *0.03* | *1.00* | *0.00* | *0.85* | *0.36* |
| 5 | 0.54 | 0.50 | 0.39 | 0.49 | 1.00 | 0.00 | 1.00 | 0.00 | 0.83 | 0.38 | 1.00 | 0.04 | 1.00 | 0.00 | 0.82 | 0.38 |
| 10 | 0.11 | 0.31 | 1.00 | 0.00 | 1.00 | 0.00 | 1.00 | 0.00 | 1.00 | 0.00 | 1.00 | 0.00 | 1.00 | 0.00 | 0.87 | 0.33 |
| *1000* | *0.17* | *0.38* | *0.68* | *0.47* | *1.00* | *0.00* | *1.00* | *0.00* | *0.94* | *0.24* | *1.00* | *0.00* | *1.00* | *0.00* | *0.83* | *0.38* |
| 5 | 0.29 | 0.45 | 0.36 | 0.48 | 1.00 | 0.00 | 1.00 | 0.00 | 0.87 | 0.33 | 1.00 | 0.00 | 1.00 | 0.00 | 0.79 | 0.41 |
| 10 | 0.06 | 0.23 | 1.00 | 0.00 | 1.00 | 0.00 | 1.00 | 0.00 | 1.00 | 0.00 | 1.00 | 0.00 | 1.00 | 0.00 | 0.87 | 0.34 |
| *5000* | *0.01* | *0.08* | *0.62* | *0.49* | *1.00* | *0.00* | *1.00* | *0.00* | *1.00* | *0.06* | *1.00* | *0.00* | *1.00* | *0.00* | *0.80* | *0.40* |
| 5 | 0.00 | 0.00 | 0.23 | 0.42 | 1.00 | 0.00 | 1.00 | 0.00 | 0.99 | 0.09 | 1.00 | 0.00 | 1.00 | 0.00 | 0.75 | 0.44 |
| 10 | 0.01 | 0.11 | 1.00 | 0.00 | 1.00 | 0.00 | 1.00 | 0.00 | 1.00 | 0.00 | 1.00 | 0.00 | 1.00 | 0.00 | 0.86 | 0.35 |
| **High Correlation** | **0.01** | **0.08** | **0.48** | **0.50** | **0.78** | **0.41** | **0.91** | **0.28** | **0.52** | **0.50** | **0.94** | **0.24** | **0.91** | **0.29** | **0.65** | **0.48** |

| | | | | | | | | | | | | | | | | | |
|---|---|---|---|---|---|---|---|---|---|---|---|---|---|---|---|---|---|
| **100** | | *0.03* | *0.16* | *0.43* | *0.50* | *0.16* | *0.36* | *0.66* | *0.47* | *0.59* | *0.49* | *0.77* | *0.42* | *0.62* | *0.48* | *0.46* | *0.50* |
| | 5 | 0.05 | 0.22 | 0.05 | 0.22 | 0.11 | 0.31 | 0.41 | 0.49 | 0.20 | 0.40 | 0.64 | 0.48 | 0.67 | 0.47 | 0.31 | 0.46 |
| | 10 | 0.00 | 0.06 | 0.81 | 0.39 | 0.20 | 0.40 | 0.91 | 0.29 | 0.97 | 0.18 | 0.90 | 0.29 | 0.58 | 0.49 | 0.62 | 0.48 |
| **500** | | *0.00* | *0.03* | *0.50* | *0.50* | *0.98* | *0.14* | *1.00* | *0.07* | *0.50* | *0.50* | *0.99* | *0.10* | *1.00* | *0.00* | *0.71* | *0.45* |
| | 5 | 0.00 | 0.04 | 0.00 | 0.00 | 0.96 | 0.20 | 0.99 | 0.10 | 0.00 | 0.04 | 0.98 | 0.15 | 1.00 | 0.00 | 0.56 | 0.50 |
| | 10 | 0.00 | 0.00 | 0.99 | 0.10 | 1.00 | 0.00 | 1.00 | 0.00 | 1.00 | 0.00 | 1.00 | 0.00 | 1.00 | 0.00 | 0.86 | 0.35 |
| **1000** | | *0.00* | *0.00* | *0.50* | *0.50* | *1.00* | *0.00* | *1.00* | *0.00* | *0.50* | *0.50* | *1.00* | *0.06* | *1.00* | *0.00* | *0.71* | *0.45* |
| | 5 | 0.00 | 0.00 | 0.00 | 0.00 | 1.00 | 0.00 | 1.00 | 0.00 | 0.00 | 0.00 | 0.99 | 0.09 | 1.00 | 0.00 | 0.57 | 0.50 |
| | 10 | 0.00 | 0.00 | 1.00 | 0.00 | 1.00 | 0.00 | 1.00 | 0.00 | 1.00 | 0.00 | 1.00 | 0.00 | 1.00 | 0.00 | 0.86 | 0.35 |
| **5000** | | *0.00* | *0.00* | *0.50* | *0.50* | *1.00* | *0.00* | *1.00* | *0.00* | *0.50* | *0.50* | *1.00* | *0.00* | *1.00* | *0.00* | *0.71* | *0.45* |
| | 5 | 0.00 | 0.00 | 0.00 | 0.00 | 1.00 | 0.00 | 1.00 | 0.00 | 0.00 | 0.00 | 1.00 | 0.00 | 1.00 | 0.00 | 0.57 | 0.49 |
| | 10 | 0.00 | 0.00 | 1.00 | 0.00 | 1.00 | 0.00 | 1.00 | 0.00 | 1.00 | 0.00 | 1.00 | 0.00 | 1.00 | 0.00 | 0.86 | 0.35 |
| | **Total** | 0.22 | 0.41 | 0.78 | 0.41 | 0.92 | 0.27 | 0.97 | 0.16 | 0.86 | 0.35 | 0.97 | 0.16 | 0.96 | 0.19 | 0.81 | 0.39 |

Fig 1 presents the mean accuracies and its 95% confidence interval by correlation (top left panel), number of items per factor (to right panel), sample size (bottom left panel) and by all conditions combined (bottom right panel). In general, the mean accuracies spread as the correlation between factors increase from zero to .7, with the Kaiser-Guttman's rule, PA and EGA presenting the highest accuracies (Fig 1, top left panel). On the other side, the mean accuracies are higher (between 90% and 100%) when the number of items increase from 5 to 10, except for the VSS (Fig 1, top right panel). As the sample size increases, the mean accuracies of BIC, eBIC, PA and EGA also increase, attaining its maximum from sample sizes of 500 on (Fig 1, bottom left panel). The Kaiser-Guttman's rule is the technique less affected by the variability in sample size. Finally, the bottom right panel of Fig 1 shows clearly that the worst scenario appears when the correlation is high (.7), the number of items is small (5 per factor) and the sample size is 100. In this case, as the sample size increases from 100 to 500, 1,000, or 5,000, BIC, eBIC, PA and EGA increase its accuracies up to 100% (Table 1).

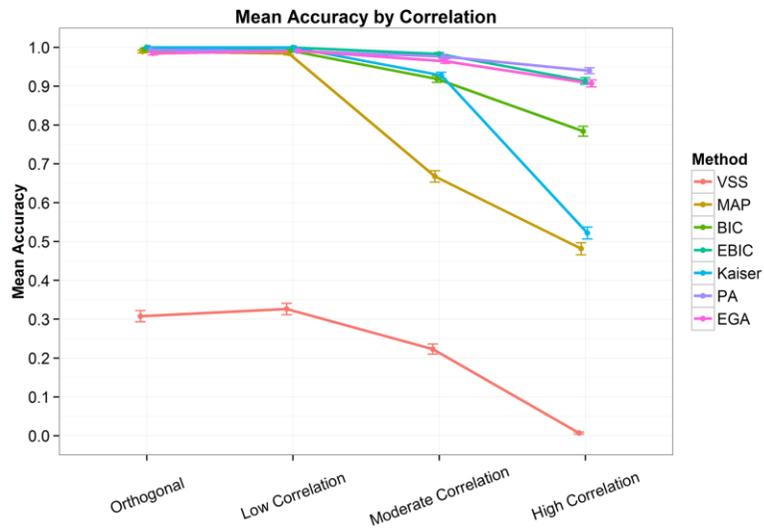
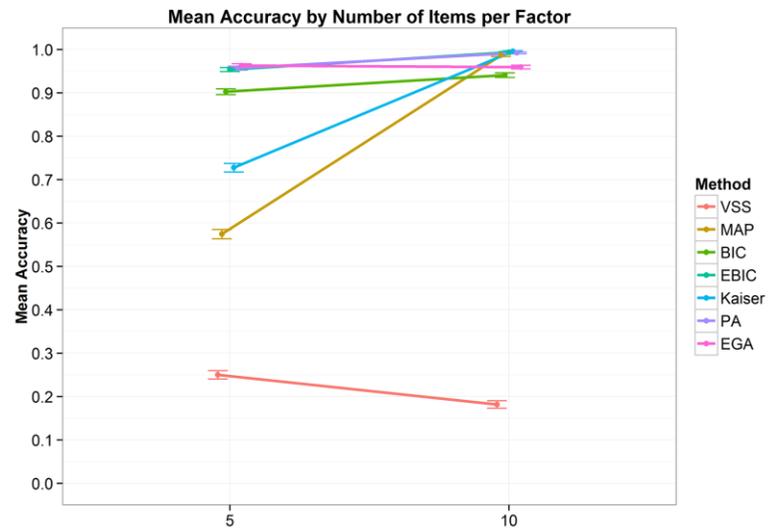
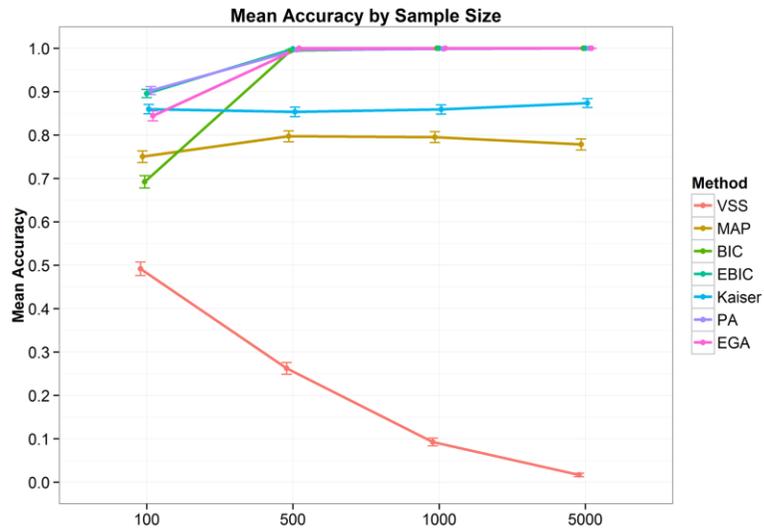
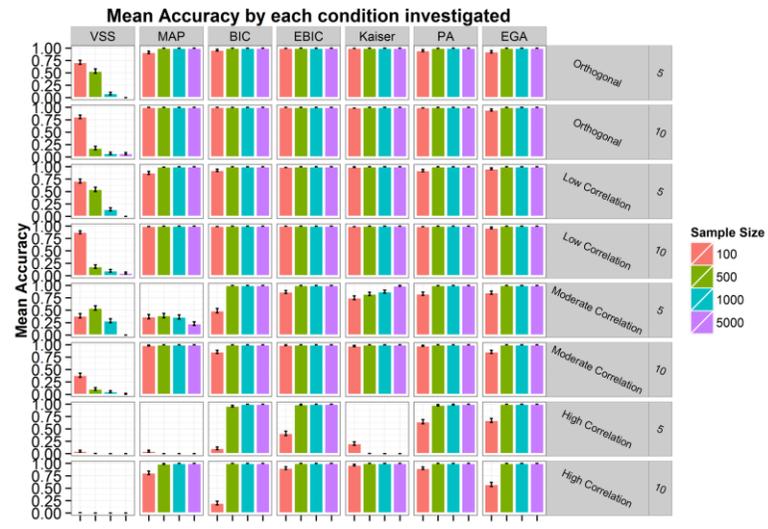

Fig 1. Mean accuracies and its 95% confidence interval by correlation (top left panel), number of items per factor (top right panel), sample size (bottom left panel) and by all conditions combined (bottom right panel) for the two-factor structure. VSS = Very Simple Structure; BIC = Bayesian Information Criteria; EBIC = Extended Bayesian Information Criteria; MAP = Minimum Average Partial procedure; Kaiser = Kaiser-Guttman eigenvalue greater than one rule; PA = Parallel Analysis; EGA = Exploratory Graph Analysis. Low correlation = .2; Moderate Correlation = .5; High Correlation = .7.

*Bias Error and Absolute Error*:

In terms of mean bias error (see Fig 2), i.e. the mean difference between the estimated and the correct number of factors, VSS presented a very high error, indicating an overestimation when the correlation between factors are orthogonal (MBE = 2.96, SD = 2.95). As the correlation increases, the mean bias error of VSS, Kaiser-Guttman rule, BIC and MAP decreases, while PA and eBIC remains relatively stable, and EGA increases the MBE from .03 (SD = .02) when the correlation is orthogonal to .19 (SD = .28) when the correlation is high (Fig 2, top left panel). From five to ten items per factor, VSS also decreases its mean bias error, while the other techniques remain stable or increases the MBE to values close to zero (Fig 2, top right panel). Considering the sample size, the highest MBE variability is found when the sample equals 100 cases (Fig 2, bottom left panel), with EGA presenting a mean bias error of .30 (SD = 1.06), while the MBE of VSS was .05 (SD = 1.54), of PA .03 (SD = .37), eBIC -.10 (SD = .31), Kaiser-Guttman -.10 (SD = .35), MAP -.25 (SD = .43), and BIC -.31 (SD = .46). The increase in sample size sharply increases the MBE of VSS. When the sample size was equal to or greater than 500, EGA, PA, BIC and eBIC presented a MBE of zero (Fig 2, bottom left panel). Analyzing all the conditions together (Fig 2, bottom right panel), it is clear that VSS is the technique presenting the more intense issue with overestimation, while MAP and the Kaiser-Guttman rule tend to underestimate the number of factors when the correlation if high and the number of items per factor is five (Fig 2, bottom right panel). In terms of mean absolute error (see Fig 3), i.e. the mean absolute difference between the estimated and the correct number of factors, the scenario is very similar to the described above.

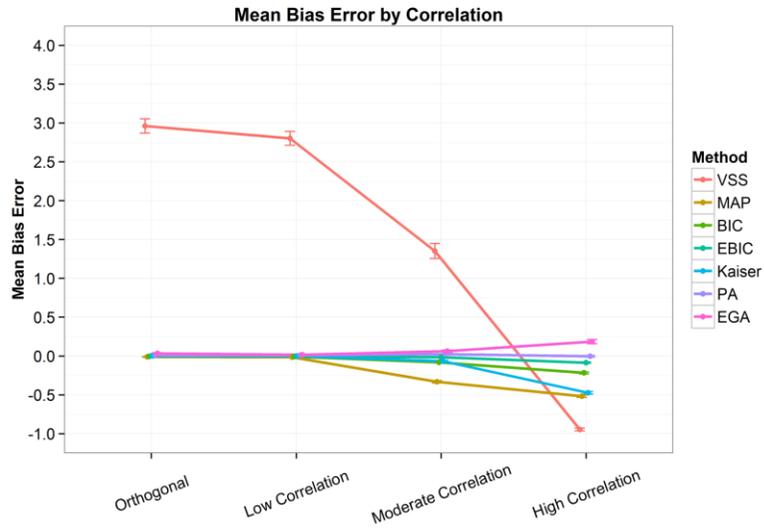
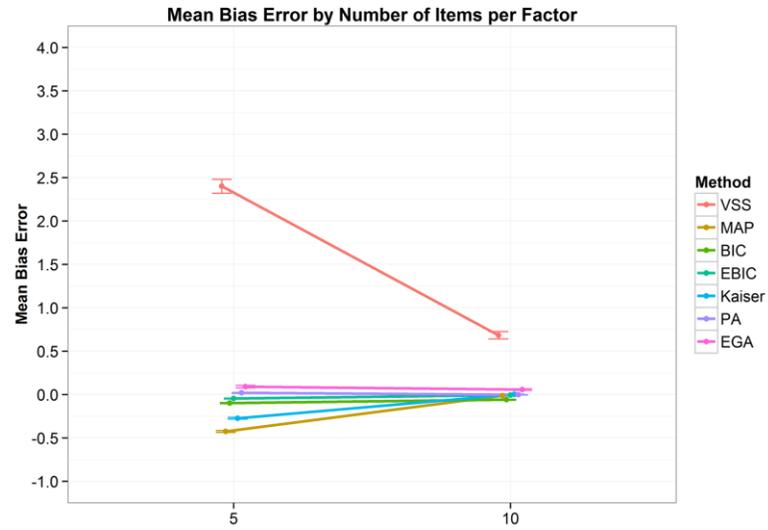
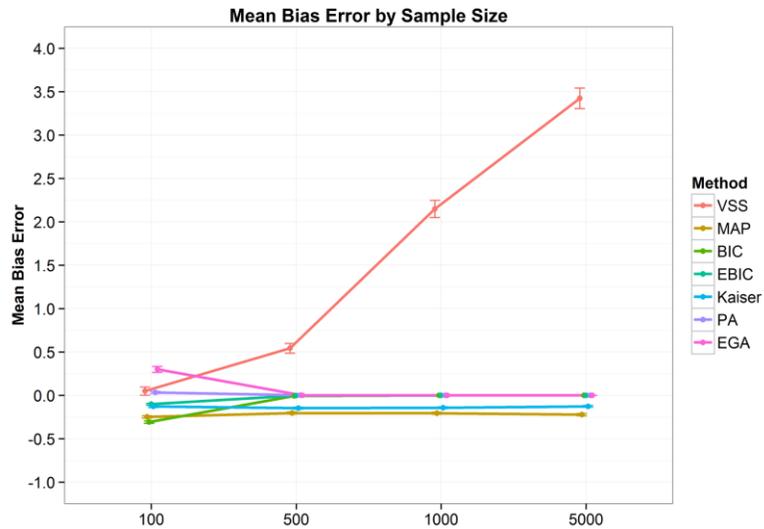
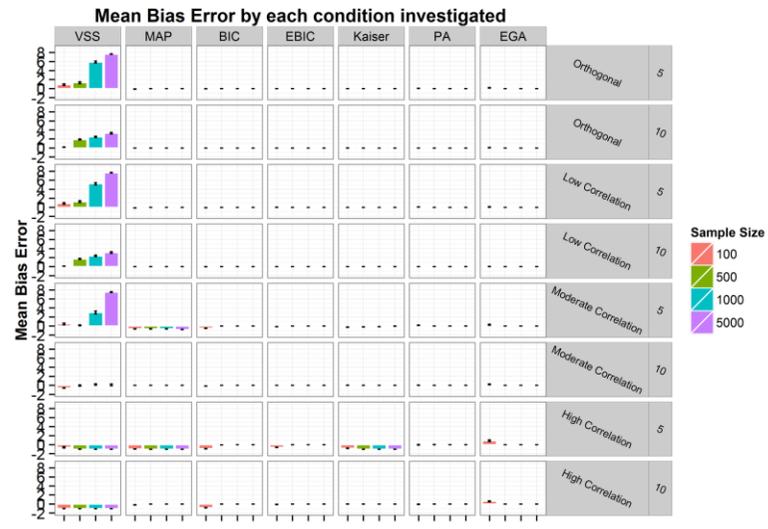

Fig 2. Mean bias error and its 95% confidence interval by correlation (top left panel), number of items per factor (top right panel), sample size (bottom left panel) and by all conditions combined (bottom right panel) for the two-factor structure. VSS = Very Simple Structure; BIC = Bayesian Information Criteria; EBIC = Extended Bayesian Information Criteria; MAP = Minimum Average Partial procedure; Kaiser = Kaiser-Guttman eigenvalue greater than one rule; PA = Parallel Analysis; EGA = Exploratory Graph Analysis. Low correlation = .2; Moderate Correlation = .5; High Correlation = .7.

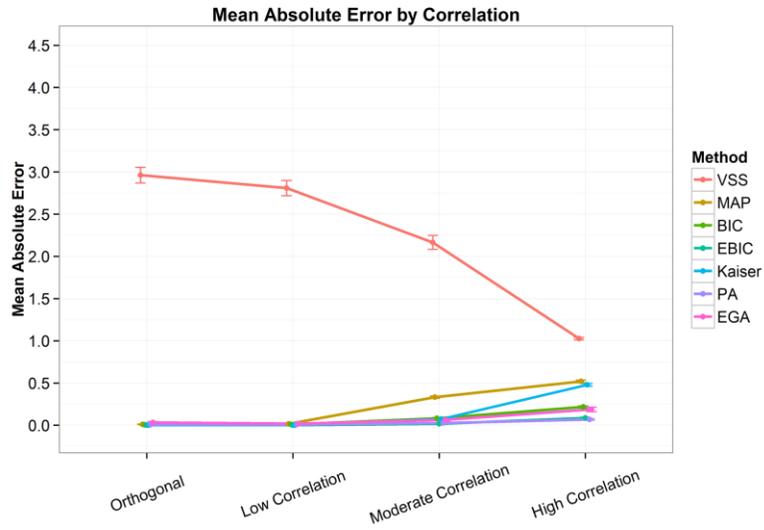
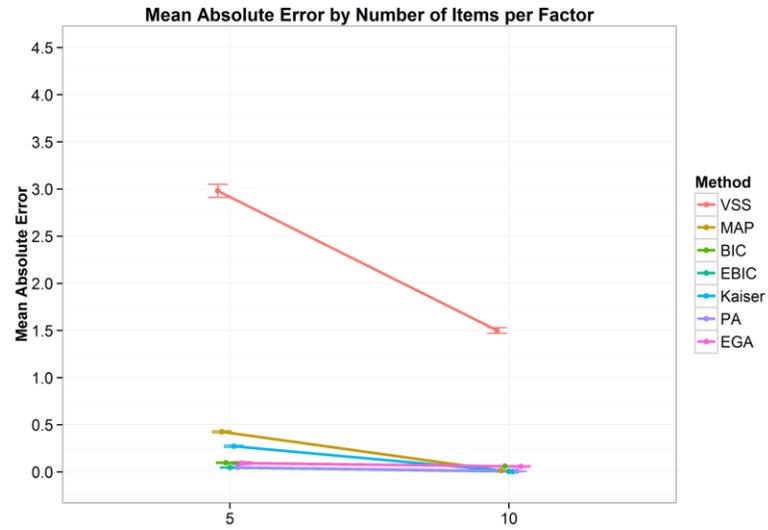
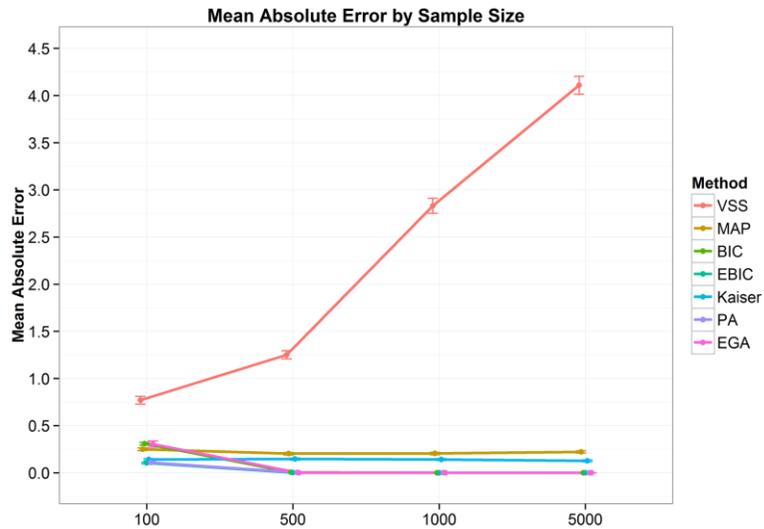
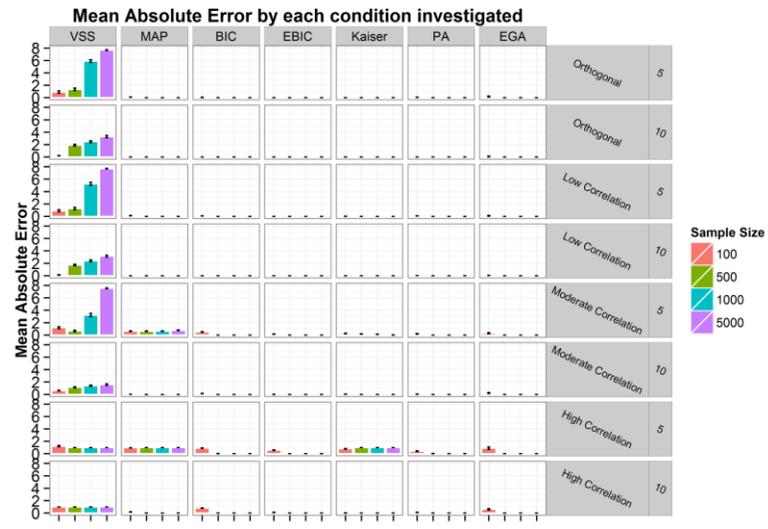

Fig 3. Mean absolute error and its 95% confidence interval by correlation (top left panel), number of items per factor (top right panel), sample size (bottom left panel) and by all conditions combined (bottom right panel) for the two-factor structure. VSS = Very Simple Structure; BIC = Bayesian Information Criteria; EBIC = Extended Bayesian Information Criteria; MAP = Minimum Average Partial procedure; Kaiser = Kaiser-Guttman eigenvalue greater than one rule; PA = Parallel Analysis; EGA = Exploratory Graph Analysis. Low correlation = .2; Moderate Correlation = .5; High Correlation = .7.

*Structure with Four Factors*

    *Accuracy*:

       Table 2 shows the mean accuracy and its standard deviation for each method, in each condition in the four factor structure. When the correlation between factors was zero (orthogonal), BIC, the Kaiser-Guttman rule, PA and EGA achieved accuracies greater than 90%, while MAP presented a mean accuracy of only 32% (SD = 47%). The increase in the sample size improved the mean accuracies, except for MAP. The same scenario appeared when the correlation between factors were low. However, when the correlation was moderate, only EGA achieved a mean accuracy greater than 90%, irrespective of the sample size, number of items per factor or sample size. In the high correlation scenario, EGA showed the higher overall accuracy (Mean = 71%, SD = 46%). However, as the sample size and the number of items per factor increased, BIC, eBIC, Kaiser-Guttman and PA were able to achieve mean accuracies greater than 90%.

**Table 2**

Mean accuracy and its standard deviation, for each method and each condition, for the four factor structure. VSS = Very Simple Structure; BIC = Bayesian Information Criteria; EBIC = Extended Bayesian Information Criteria; MAP = Minimum Average Partial procedure; Kaiser = Kaiser-Guttman eigenvalue greater than one rule; PARAN = Parallel Analysis; EGA = Exploratory Graph Analysis. Low correlation = .2; Moderate Correlation = .5; High Correlation = .7. The rows show the aggregate mean and standard deviation for each level of correlation (bold), sample size (bold and italicized) and number of items per factor (non-italicized).

| | | VSS | | MAP | | BIC | | EBIC | | Kaiser | | PA | | EGA | | Total | |
|---|---|---|---|---|---|---|---|---|---|---|---|---|---|---|---|---|---|
| | | Mean | SD | Mean | SD | Mean | SD | Mean | SD | Mean | SD | Mean | SD | __Mean__ | __SD__ | Mean | SD |
| **Orthogonal** | | **0.79** | **0.41** | **0.32** | **0.47** | **0.95** | **0.23** | **0.86** | **0.35** | **0.99** | **0.11** | **1.00** | **0.07** | __**0.95**__ | __**0.21**__ | **0.83** | **0.37** |
| *100* | | *0.72* | *0.45* | *0.54* | *0.50* | *0.79* | *0.41* | *0.45* | *0.50* | *0.95* | *0.22* | *0.98* | *0.13* | __*0.75*__ | __*0.43*__ | *0.74* | *0.44* |
| | 5 | 0.89 | 0.32 | 0.57 | 0.50 | 0.68 | 0.47 | 0.26 | 0.44 | 0.98 | 0.15 | 0.97 | 0.18 | __0.68__ | __0.47__ | 0.72 | 0.45 |
| | 10 | 0.55 | 0.50 | 0.51 | 0.50 | 0.89 | 0.31 | 0.63 | 0.48 | 0.92 | 0.27 | 1.00 | 0.00 | __0.93__ | __0.25__ | 0.76 | 0.43 |
| *500* | | *0.81* | *0.39* | *0.26* | *0.44* | *1.00* | *0.00* | *1.00* | *0.00* | *1.00* | *0.00* | *1.00* | *0.00* | __*1.00*__ | __*0.00*__ | *0.87* | *0.34* |
| | 5 | 0.81 | 0.39 | 0.30 | 0.46 | 1.00 | 0.00 | 1.00 | 0.00 | 1.00 | 0.00 | 1.00 | 0.00 | __1.00__ | __0.00__ | 0.87 | 0.33 |
| | 10 | 0.80 | 0.40 | 0.21 | 0.40 | 1.00 | 0.00 | 1.00 | 0.00 | 1.00 | 0.00 | 1.00 | 0.00 | __1.00__ | __0.00__ | 0.86 | 0.35 |
| *1000* | | *0.81* | *0.39* | *0.25* | *0.43* | *1.00* | *0.00* | *1.00* | *0.00* | *1.00* | *0.00* | *1.00* | *0.00* | __*1.00*__ | __*0.00*__ | *0.87* | *0.34* |
| | 5 | 0.83 | 0.38 | 0.30 | 0.46 | 1.00 | 0.00 | 1.00 | 0.00 | 1.00 | 0.00 | 1.00 | 0.00 | __1.00__ | __0.00__ | 0.87 | 0.33 |
| | 10 | 0.80 | 0.40 | 0.20 | 0.40 | 1.00 | 0.00 | 1.00 | 0.00 | 1.00 | 0.00 | 1.00 | 0.00 | __1.00__ | __0.00__ | 0.86 | 0.35 |
| *5000* | | *0.81* | *0.39* | *0.23* | *0.42* | *1.00* | *0.00* | *1.00* | *0.00* | *1.00* | *0.00* | *1.00* | *0.00* | __*1.00*__ | __*0.00*__ | *0.86* | *0.34* |
| | 5 | 0.82 | 0.38 | 0.26 | 0.44 | 1.00 | 0.00 | 1.00 | 0.00 | 1.00 | 0.00 | 1.00 | 0.00 | __1.00__ | __0.00__ | 0.87 | 0.34 |
| | 10 | 0.80 | 0.40 | 0.20 | 0.40 | 1.00 | 0.00 | 1.00 | 0.00 | 1.00 | 0.00 | 1.00 | 0.00 | __1.00__ | __0.00__ | 0.86 | 0.35 |
| **Low Correlation** | | **0.76** | **0.43** | **0.31** | **0.46** | **0.91** | **0.29** | **0.82** | **0.39** | **0.97** | **0.18** | **0.99** | **0.10** | __**0.97**__ | __**0.17**__ | **0.82** | **0.39** |
| *100* | | *0.62* | *0.48* | *0.51* | *0.50* | *0.64* | *0.48* | *0.27* | *0.45* | *0.87* | *0.34* | *0.96* | *0.19* | __*0.82*__ | __*0.38*__ | *0.66* | *0.47* |
| | 5 | 0.82 | 0.38 | 0.50 | 0.50 | 0.48 | 0.50 | 0.19 | 0.40 | 0.85 | 0.36 | 0.93 | 0.26 | __0.79__ | __0.41__ | 0.65 | 0.48 |
| | 10 | 0.42 | 0.49 | 0.51 | 0.50 | 0.80 | 0.40 | 0.35 | 0.48 | 0.89 | 0.31 | 1.00 | 0.04 | __0.97__ | __0.16__ | 0.67 | 0.47 |
| *500* | | *0.81* | *0.39* | *0.25* | *0.43* | *1.00* | *0.03* | *1.00* | *0.00* | *1.00* | *0.00* | *1.00* | *0.00* | __*1.00*__ | __*0.00*__ | *0.87* | *0.34* |
| | 5 | 0.82 | 0.39 | 0.30 | 0.46 | 1.00 | 0.04 | 1.00 | 0.00 | 1.00 | 0.00 | 1.00 | 0.00 | __1.00__ | __0.00__ | 0.87 | 0.33 |
| | 10 | 0.80 | 0.40 | 0.20 | 0.40 | 1.00 | 0.00 | 1.00 | 0.00 | 1.00 | 0.00 | 1.00 | 0.00 | __1.00__ | __0.00__ | 0.86 | 0.35 |
| *1000* | | *0.81* | *0.39* | *0.24* | *0.43* | *1.00* | *0.00* | *1.00* | *0.00* | *1.00* | *0.00* | *1.00* | *0.00* | __*1.00*__ | __*0.00*__ | *0.86* | *0.34* |
| | 5 | 0.82 | 0.38 | 0.28 | 0.45 | 1.00 | 0.00 | 1.00 | 0.00 | 1.00 | 0.00 | 1.00 | 0.00 | __1.00__ | __0.00__ | 0.87 | 0.34 |
| | 10 | 0.80 | 0.40 | 0.20 | 0.40 | 1.00 | 0.00 | 1.00 | 0.00 | 1.00 | 0.00 | 1.00 | 0.00 | __1.00__ | __0.00__ | 0.86 | 0.35 |

| | | | | | | | | | | | | | | | | |
|---|---|---|---|---|---|---|---|---|---|---|---|---|---|---|---|---|
| *5000* | *0.81* | *0.40* | *0.23* | *0.42* | *1.00* | *0.00* | *1.00* | *0.00* | *1.00* | *0.00* | *1.00* | *0.00* | *__1.00__* | *__0.00__* | *0.86* | *0.34* |
| 5 | 0.81 | 0.39 | 0.26 | 0.44 | 1.00 | 0.00 | 1.00 | 0.00 | 1.00 | 0.00 | 1.00 | 0.00 | __1.00__ | __0.00__ | 0.87 | 0.34 |
| 10 | 0.80 | 0.40 | 0.20 | 0.40 | 1.00 | 0.00 | 1.00 | 0.00 | 1.00 | 0.00 | 1.00 | 0.00 | __1.00__ | __0.00__ | 0.86 | 0.35 |
| **Moderate Correlation** | **0.33** | **0.47** | **0.13** | **0.33** | **0.51** | **0.50** | **0.63** | **0.48** | **0.70** | **0.46** | **0.83** | **0.37** | **__0.93__** | **__0.25__** | **0.57** | **0.49** |
| *100* | *0.09* | *0.29* | *0.13* | *0.34* | *0.36* | *0.48* | *0.10* | *0.29* | *0.39* | *0.49* | *0.55* | *0.50* | *__0.56__* | *__0.50__* | *0.29* | *0.46* |
| 5 | 0.04 | 0.21 | 0.07 | 0.25 | 0.01 | 0.08 | 0.00 | 0.00 | 0.03 | 0.17 | 0.18 | 0.38 | __0.50__ | __0.50__ | 0.11 | 0.31 |
| 10 | 0.14 | 0.34 | 0.20 | 0.40 | 0.72 | 0.45 | 0.19 | 0.39 | 0.75 | 0.44 | 0.93 | 0.26 | __0.97__ | __0.16__ | 0.50 | 0.50 |
| *500* | *0.41* | *0.49* | *0.13* | *0.34* | *0.51* | *0.50* | *0.56* | *0.50* | *0.64* | *0.48* | *0.81* | *0.40* | *__1.00__* | *__0.04__* | *0.58* | *0.49* |
| 5 | 0.01 | 0.10 | 0.07 | 0.25 | 0.01 | 0.10 | 0.11 | 0.31 | 0.28 | 0.45 | 0.61 | 0.49 | __1.00__ | __0.06__ | 0.30 | 0.46 |
| 10 | 0.80 | 0.40 | 0.20 | 0.40 | 1.00 | 0.00 | 1.00 | 0.00 | 1.00 | 0.00 | 1.00 | 0.00 | __1.00__ | __0.00__ | 0.86 | 0.35 |
| *1000* | *0.41* | *0.49* | *0.12* | *0.32* | *0.57* | *0.49* | *0.88* | *0.33* | *0.87* | *0.33* | *0.97* | *0.18* | *__1.00__* | *__0.00__* | *0.69* | *0.46* |
| 5 | 0.01 | 0.10 | 0.04 | 0.19 | 0.15 | 0.35 | 0.75 | 0.43 | 0.74 | 0.44 | 0.93 | 0.26 | __1.00__ | __0.00__ | 0.52 | 0.50 |
| 10 | 0.80 | 0.40 | 0.20 | 0.40 | 1.00 | 0.00 | 1.00 | 0.00 | 1.00 | 0.00 | 1.00 | 0.00 | __1.00__ | __0.00__ | 0.86 | 0.35 |
| *5000* | *0.41* | *0.49* | *0.12* | *0.32* | *0.60* | *0.49* | *1.00* | *0.00* | *0.90* | *0.30* | *1.00* | *0.00* | *__1.00__* | *__0.00__* | *0.72* | *0.45* |
| 5 | 0.01 | 0.12 | 0.03 | 0.18 | 0.20 | 0.40 | 1.00 | 0.00 | 0.80 | 0.40 | 1.00 | 0.00 | __1.00__ | __0.00__ | 0.58 | 0.49 |
| 10 | 0.80 | 0.40 | 0.20 | 0.40 | 1.00 | 0.00 | 1.00 | 0.00 | 1.00 | 0.00 | 1.00 | 0.00 | __1.00__ | __0.00__ | 0.86 | 0.35 |
| **High Correlation** | **0.32** | **0.47** | **0.08** | **0.27** | **0.39** | **0.49** | **0.35** | **0.48** | **0.39** | **0.49** | **0.40** | **0.49** | **__0.71__** | **__0.46__** | **0.37** | **0.48** |
| *100* | *0.07* | *0.26* | *0.03* | *0.18* | *0.14* | *0.35* | *0.01* | *0.10* | *0.07* | *0.25* | *0.10* | *0.30* | *__0.27__* | *__0.44__* | *0.09* | *0.28* |
| 5 | 0.00 | 0.00 | 0.00 | 0.00 | 0.00 | 0.00 | 0.00 | 0.00 | 0.00 | 0.00 | 0.00 | 0.02 | __0.15__ | __0.36__ | 0.02 | 0.15 |
| 10 | 0.15 | 0.35 | 0.06 | 0.24 | 0.28 | 0.45 | 0.02 | 0.15 | 0.13 | 0.34 | 0.17 | 0.38 | __0.64__ | __0.48__ | 0.16 | 0.36 |
| *500* | *0.40* | *0.49* | *0.09* | *0.29* | *0.44* | *0.50* | *0.38* | *0.49* | *0.50* | *0.50* | *0.50* | *0.50* | *__0.61__* | *__0.49__* | *0.42* | *0.49* |
| 5 | 0.00 | 0.00 | 0.00 | 0.00 | 0.00 | 0.00 | 0.00 | 0.00 | 0.00 | 0.00 | 0.00 | 0.04 | __0.22__ | __0.42__ | 0.03 | 0.18 |
| 10 | 0.79 | 0.41 | 0.19 | 0.39 | 0.87 | 0.33 | 0.76 | 0.43 | 1.00 | 0.00 | 1.00 | 0.00 | __1.00__ | __0.00__ | 0.80 | 0.40 |
| *1000* | *0.40* | *0.49* | *0.10* | *0.30* | *0.50* | *0.50* | *0.50* | *0.50* | *0.50* | *0.50* | *0.50* | *0.50* | *__0.77__* | *__0.42__* | *0.47* | *0.50* |
| 5 | 0.00 | 0.00 | 0.00 | 0.00 | 0.00 | 0.00 | 0.00 | 0.00 | 0.00 | 0.00 | 0.00 | 0.00 | __0.53__ | __0.50__ | 0.08 | 0.27 |
| 10 | 0.80 | 0.40 | 0.20 | 0.40 | 1.00 | 0.06 | 1.00 | 0.00 | 1.00 | 0.00 | 1.00 | 0.00 | __1.00__ | __0.00__ | 0.86 | 0.35 |
| *5000* | *0.40* | *0.49* | *0.10* | *0.30* | *0.50* | *0.50* | *0.50* | *0.50* | *0.50* | *0.50* | *0.50* | *0.50* | *__1.00__* | *__0.00__* | *0.50* | *0.50* |
| 5 | 0.00 | 0.00 | 0.00 | 0.00 | 0.00 | 0.00 | 0.00 | 0.00 | 0.00 | 0.00 | 0.00 | 0.00 | __1.00__ | __0.00__ | 0.14 | 0.35 |
| 10 | 0.80 | 0.40 | 0.20 | 0.40 | 1.00 | 0.00 | 1.00 | 0.00 | 1.00 | 0.00 | 1.00 | 0.00 | __1.00__ | __0.00__ | 0.86 | 0.35 |
| **Total** | **0.55** | **0.50** | **0.21** | **0.41** | **0.69** | **0.46** | **0.66** | **0.47** | **0.76** | **0.43** | **0.80** | **0.40** | **__0.89__** | **__0.31__** | **0.65** | **0.48** |

Fig 4 presents the mean accuracies and its 95% confidence interval by correlation (top left panel), number of items per factor (to right panel), sample size (bottom left panel) and by all conditions combined (bottom right panel) in the four factor structure. In general, the mean accuracies decrease as the correlation between factors increases from zero to .7, with EGA presenting the highest mean accuracy (Fig 4, top left panel). On the other hand, the mean accuracies increase when the number of items goes from 5 to 10 (Fig 4, top right panel) and with the increase of the sample size (Fig 4, bottom left panel), except for MAP, whose accuracy is inversely related to sample size. Finally, the bottom right panel of Fig 4 shows, again, that the worst scenario appears when the correlation between factors is high (.7) and the number of items is small (5 per factor). In this case, only EGA was able to correctly estimate the number of dimensions, presenting a mean accuracy of 100% for a sample sizes of 5,000 (Fig 4, bottom right panel). However, the increase in the number of items per factor, from five to ten, sharply increments the mean accuracy of the methods (Fig

4, bottom right panel).

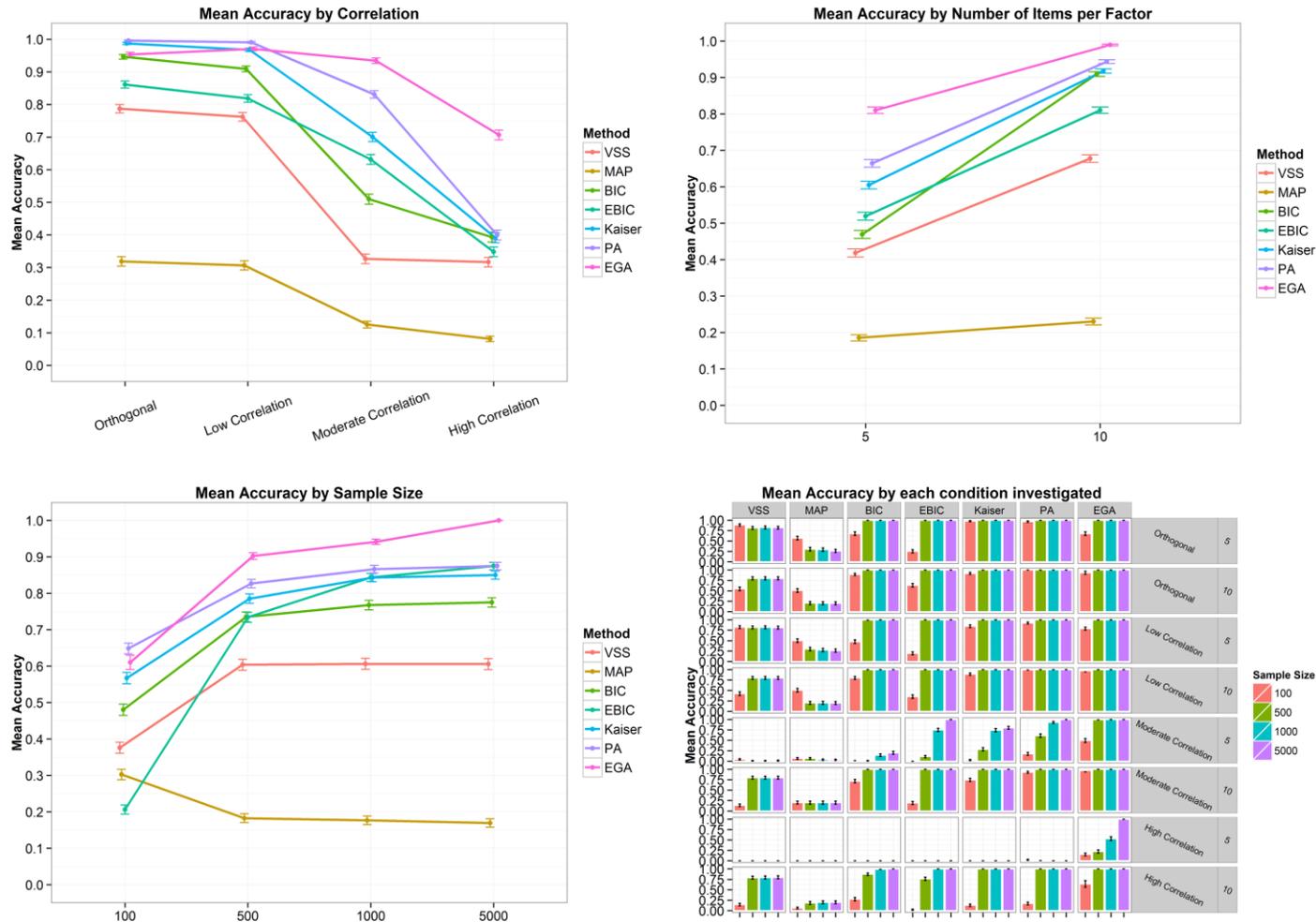

Fig 4. Mean accuracy and its 95% confidence interval by correlation (top left panel), number of items per factor (top right panel), sample size (bottom left panel) and by all conditions combined (bottom right panel) for the four-factor structure. VSS = Very Simple Structure; BIC = Bayesian Information Criteria; EBIC = Extended Bayesian Information Criteria; MAP = Minimum Average Partial procedure; Kaiser = Kaiser-Guttman eigenvalue greater than one rule; PA = Parallel Analysis; EGA = Exploratory Graph Analysis. Low correlation = .2; Moderate Correlation = .5; High Correlation = .7.

*Bias Error and Absolute Error*:

In terms of mean bias error, Fig 5 shows that VSS overestimated the number of dimensions when the correlation between factors was orthogonal or low. When the correlation between factors was moderate, MAP, BIC, eBIC, VSS, Kaiser-Guttman and PA underestimated the number of dimensions. All techniques presented a mean bias error lower than zero, indicating a tendency to underestimate the number of factors in the high correlation scenario (Fig 5, top left panel). The top right panel of Fig 5 also shows a very clear tendency: except for EGA, all the methods increased the mean bias error with the increase in number of items increases per factor. The sample size also affects the mean bias error (Fig 5, bottom left panel). When the sample size was 100, BIC, VSS, MAP, eBIC and PA presented the lowest mean bias error. As sample size increase, the mean bias error of the techniques tends to be closer to zero, except for the VSS, since the increase in the sample size implies an increase in its overestimation. Finally, the bottom right panel of Fig 5 shows what happens when the correlation between factors is high and the number of items is five: the methods tends to underestimate the number of dimensions. In terms of mean absolute error (see Fig 6), i.e. the mean absolute difference between the estimated and the correct number of factors, the scenario is very similar to the described above. In general, the absolute error increased as the correlation between factors became stronger and decreased when the number of items went from five to ten and when the sample size increased (except for VSS). EGA was the only technique to present a mean absolute error of zero for a sample size of 5,000.

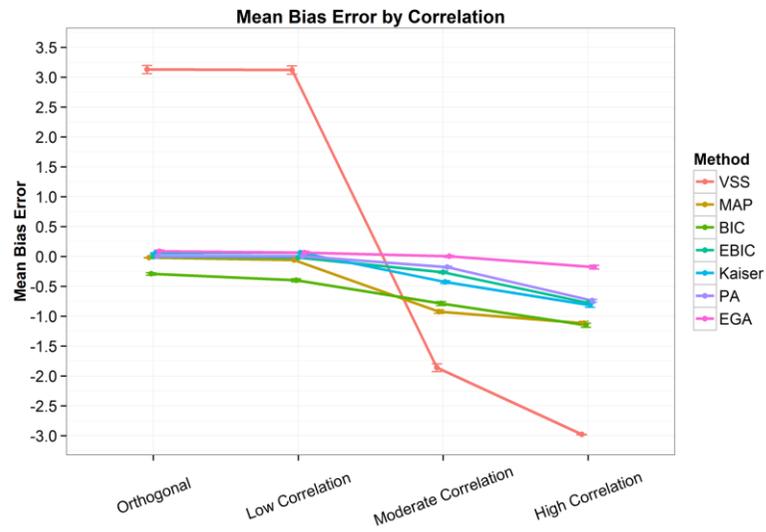
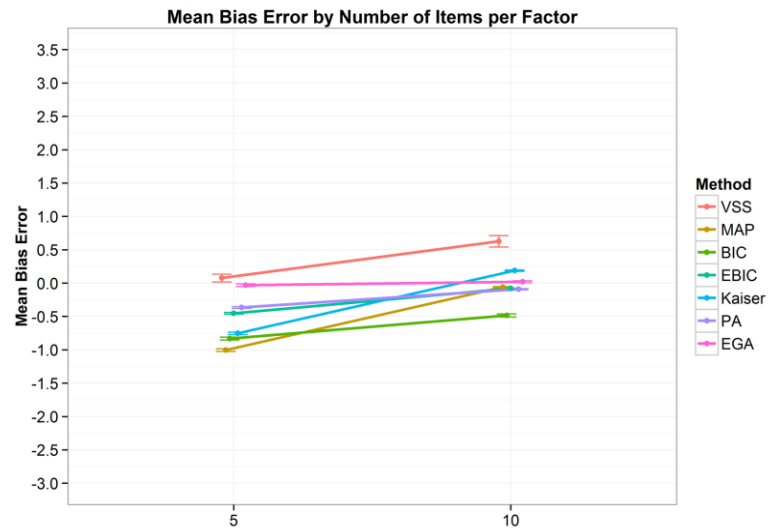
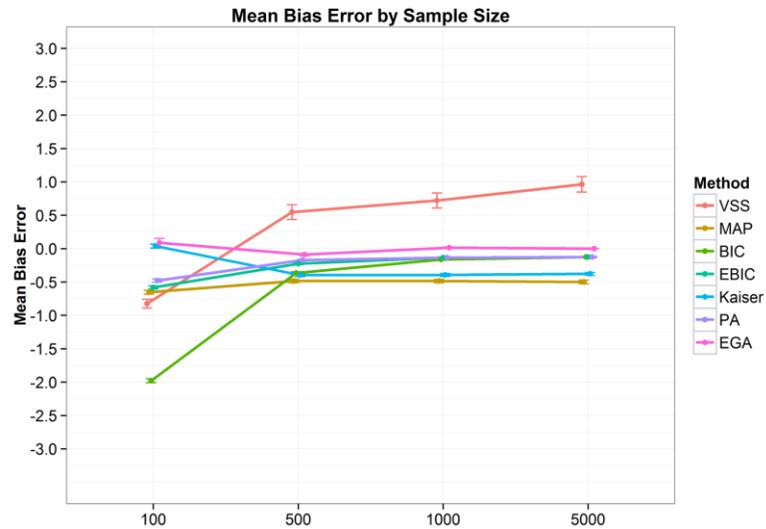
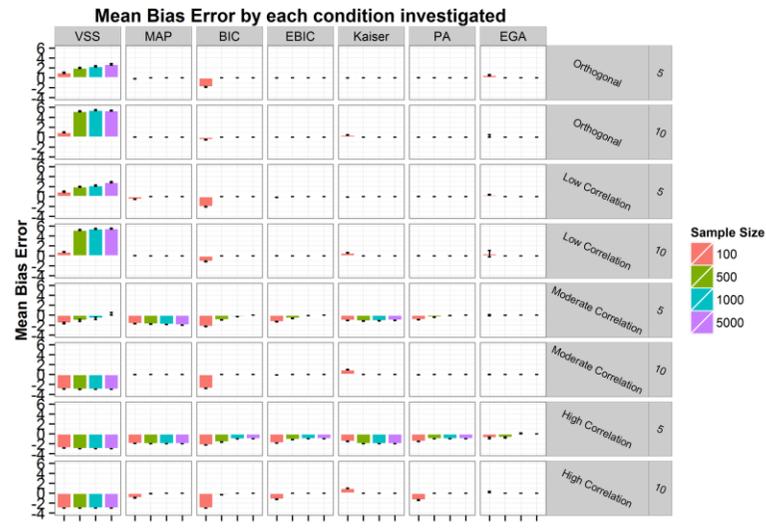

Fig 5. Mean bias error and its 95% confidence interval by correlation (top left panel), number of items per factor (top right panel), sample size (bottom left panel) and by all conditions combined (bottom right panel) for the four-factor structure. VSS = Very Simple Structure; BIC = Bayesian Information Criteria; EBIC = Extended Bayesian Information Criteria; MAP = Minimum Average Partial procedure; Kaiser = Kaiser-Guttman eigenvalue greater than one rule; PA = Parallel Analysis; EGA = Exploratory Graph Analysis. Low correlation = .2; Moderate Correlation = .5; High Correlation = .7.

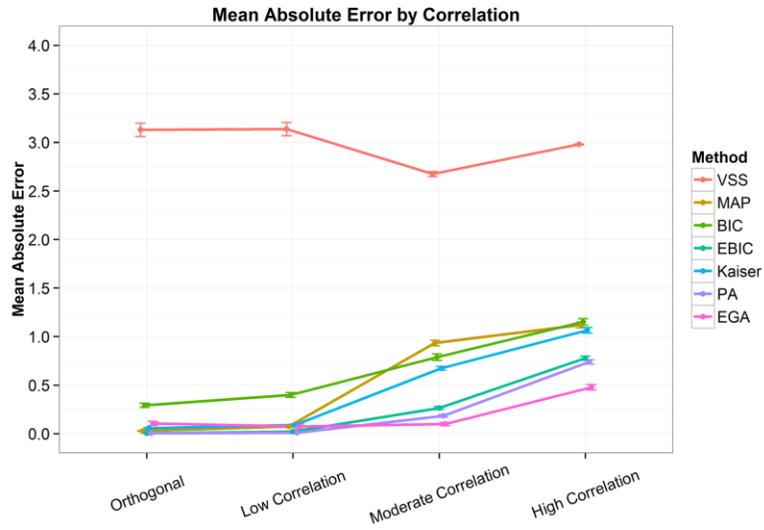
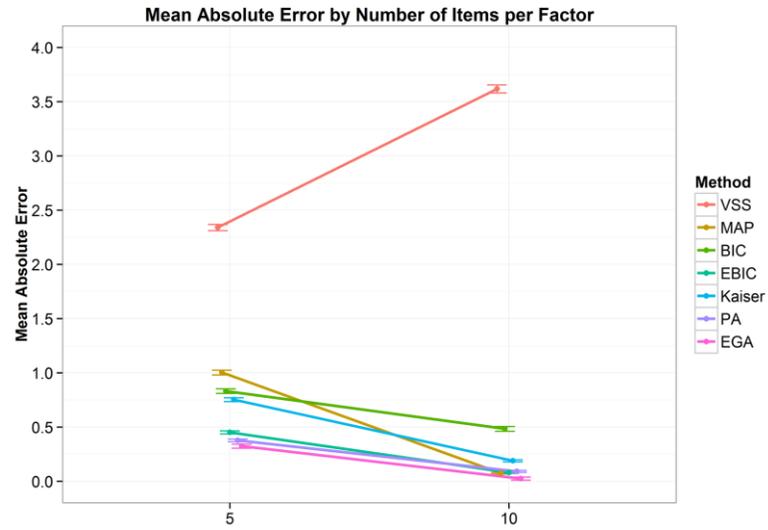
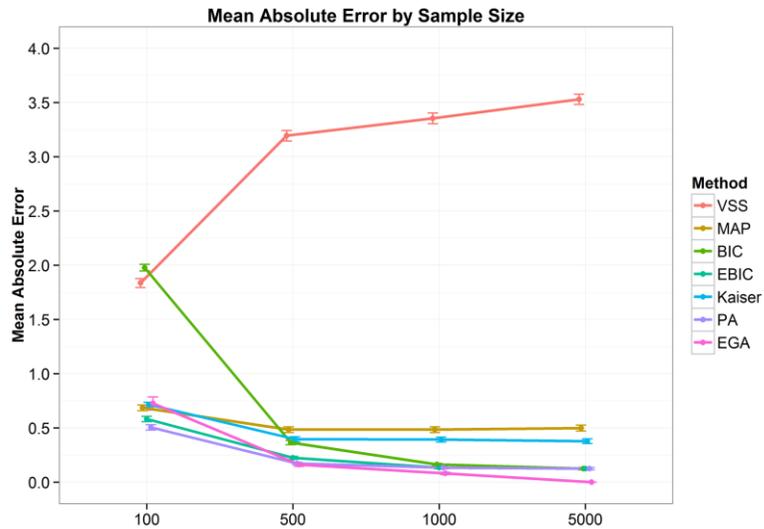
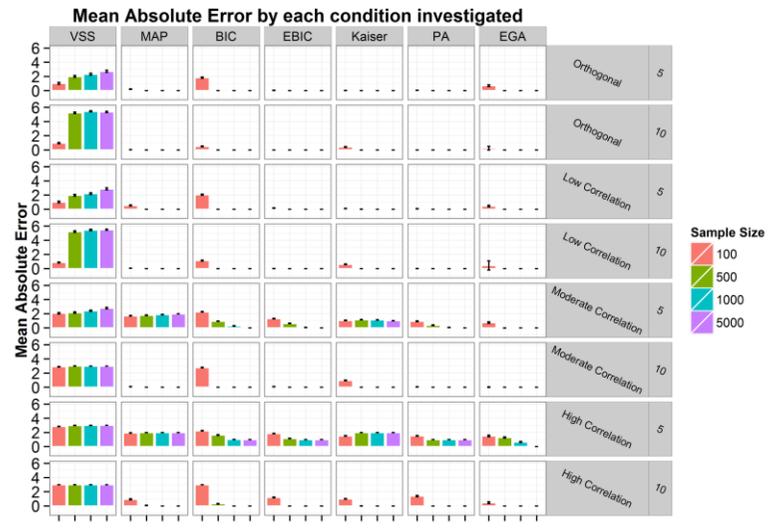

Fig 6. Mean absolute error and its 95% confidence interval by correlation (top left panel), number of items per factor (top right panel), sample size (bottom left panel) and by all conditions combined (bottom right panel) for the four-factor structure. VSS = Very Simple Structure; BIC = Bayesian Information Criteria; EBIC = Extended Bayesian Information Criteria; MAP = Minimum Average Partial procedure; Kaiser = Kaiser-Guttman eigenvalue greater than one rule; PA = Parallel Analysis; EGA = Exploratory Graph Analysis. Low correlation = .2; Moderate Correlation = .5; High Correlation = .7.

*High Order Interactions*

The final analysis aimed to verify how each condition investigated, and their combinations, impacted the accuracy to identify the correct number of dimensions for each technique used. In order to do it, an analysis of variance (ANOVA) were performed for each technique, with the accuracy as the dependent variable and the correlation between factors, sample size, number of items per factor and number of factors as the independent variables. Only the partial eta squared effect size will be reported, since the goal is to verify the magnitude of the difference between groups of conditions, in each technique. Partial eta squared values equals to or greater than .14 can be considered large effect sizes [61]. The VSS technique presented a large effect size for correlation, number of factors and for the two-way interaction of sample size X number of factors. The MAP method presented a large effect size for correlation, number of factors and for the two-way interaction of correlation X items per factor. BIC and eBIC, on the other hand, presented large effect sizes for correlation, sample size, items per factor and number of factors. BIC also presented large effect sizes for every two-way interactions involving correlation, plus the two-way interaction of items per factor X number of factors and the three-way interaction of correlation X items per factor X number of factors. The eBIC technique, on the other hand, also presented large effect sizes for correlation X items per factor, correlation X number of factors, sample size X number of factors and for the four-way interaction of correlation X sample size X items per factor X number of factors. By its turn, the Kaiser-Guttman rule presented large effect sizes for correlation, items per factor, and correlation X items per factor. Parallel analysis presented large effect sizes for all isolate conditions, plus the two-way interactions of correlation X items per factor, correlation X number of factors, as well as for the three-way interaction of correlation X items per factor X number of factors and the four-way interaction of correlation X sample size X items per factor X number of factors. Finally, EGA only presented a large effect size for the sample size, being the technique whose accuracy was least affected by the conditions investigated in this paper.

**Table 2**

ANOVA's Partial eta squared effect sizes. VSS = Very Simple Structure; BIC = Bayesian Information Criteria; EBIC = Extended Bayesian Information Criteria; MAP = Minimum Average Partial procedure; Kaiser = Kaiser-Guttman eigenvalue greater than one rule; PA = Parallel Analysis; EGA = Exploratory Graph Analysis. In bold and underlined are the large effect sizes [61].

| | VSS | MAP | BIC | EBIC | Kaiser | PA | EGA |
|---|---|---|---|---|---|---|---|
| Correlation | **0.21** | **0.22** | **0.42** | **0.32** | **0.61** | **0.44** | 0.12 |
| Sample Size | 0.03 | 0.00 | **0.31** | **0.45** | 0.10 | **0.17** | **0.18** |
| Items per Factor | 0.02 | 0.13 | **0.28** | **0.19** | **0.42** | **0.22** | 0.02 |
| Number of Factors | **0.20** | **0.47** | **0.27** | **0.44** | 0.08 | **0.25** | 0.04 |
| Correlation X Sample Size | 0.05 | 0.01 | **0.14** | 0.05 | 0.07 | 0.12 | 0.08 |
| Correlation X Items per Factor | 0.07 | **0.15** | **0.21** | **0.17** | **0.51** | **0.25** | 0.03 |
| Sample Size X Items per Factor | 0.02 | 0.00 | 0.00 | 0.01 | 0.02 | 0.01 | 0.01 |
| Correlation X Number of Factors | 0.05 | 0.03 | **0.16** | **0.19** | 0.06 | **0.36** | 0.05 |
| Sample Size X Number of Factors | **0.15** | 0.01 | 0.00 | **0.30** | 0.10 | 0.03 | 0.03 |
| Items per Factor X Number of Factors | 0.06 | 0.08 | **0.22** | 0.12 | 0.00 | 0.14 | 0.03 |
| Correlation X Sample Size X Items per Factor | 0.01 | 0.01 | 0.05 | 0.08 | 0.13 | 0.12 | 0.03 |
| Correlation X Sample Size X Number of Factors | 0.02 | 0.01 | 0.12 | 0.11 | 0.06 | 0.04 | 0.03 |
| Correlation X Items per Factor X Number of Factors | 0.08 | 0.05 | **0.17** | 0.08 | 0.07 | **0.19** | 0.05 |
| Sample Size X Items per Factor X Number of Factors | 0.05 | 0.00 | 0.02 | 0.04 | 0.01 | 0.02 | 0.02 |
| Correlation X Sample Size X Items per Factor X Number of Factors | 0.01 | 0.00 | 0.06 | **0.19** | 0.05 | **0.16** | 0.03 |

### Using EGA in real dataset

The dataset we are using in this section was published by Golino and Gomes [62]. It presents data from 1,803 Brazilians (52.5% female) with age varying from 5 to 85 years (M = 15.75; SD = 12.21) that answered to the Inductive Reasoning Developmental Test – IRDT (3rd version) [62], a pencil-and-paper instrument with 56 items designed to assess developmentally sequenced and hierarchically organized inductive reasoning. The dataset can be downloaded for reproducible purposes in the following link: https://figshare.com/articles/TDRI_dataset_csv/3142321. The sequence of IRDT items was constructed to measure seven developmental stages based on the Model of Hierarchical Complexity [63, 64] and on Fischer's Dynamic Skill Theory [65, 66], two neo-Piagetian theories of development. Golino and Gomes [62] showed that two structures can be used to describe the IRDT items. The first one is a seven correlated factors model [$\chi^2$ (1463) = 764,28; p = 0,00; CFI = 1,00; RMSEA = 0,00; NFI = 0,99; NNFI = 1,00], in which each factor represents one stage and explains a group of eight items (Fig 7). The other is a bifactor (Schmid-Leiman) model with seven specific first order factors (Fig 8), each one representing one stage and explaining a group of eight items, plus a general first order factor directly explaining the IRDT's 56 items [$\chi^2$ (1428) = 2768,36; p = 0,00; CFI = 0,98; RMSEA = 0,04; NFI = 0,95; NNFI = 0,98]. The authors showed that both models are not significantly different, via the Satorra and Bentler's [67] scaled chi-square test [$\Delta\chi^2$ = -99.87; $\Delta$DF = 35; p = 1]. Figs 7 and 8 shows the standardized factor loadings and correlations of both models, and were created using *semPlot* [68].

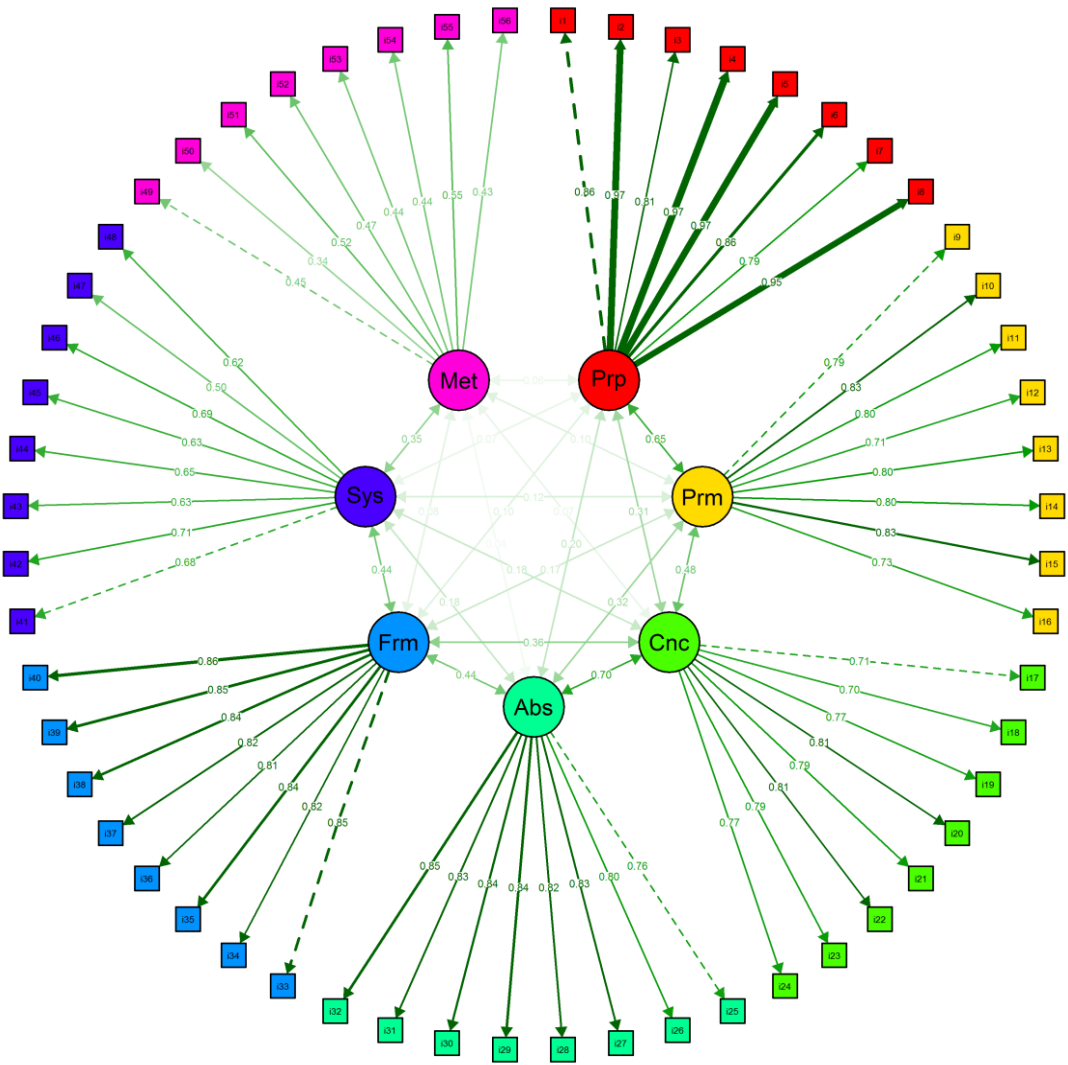

Fig 7. Inductive Reasoning Developmental Test [62] seven correlated factors model. The factors correspond to the stages the instrument intended to measure: Prp = Pre-Operational; Prm = Primary; Cnc = Concrete; Abs = Abstract; Frm = Formal; Sys = Systematic; Met = Metasystematic.

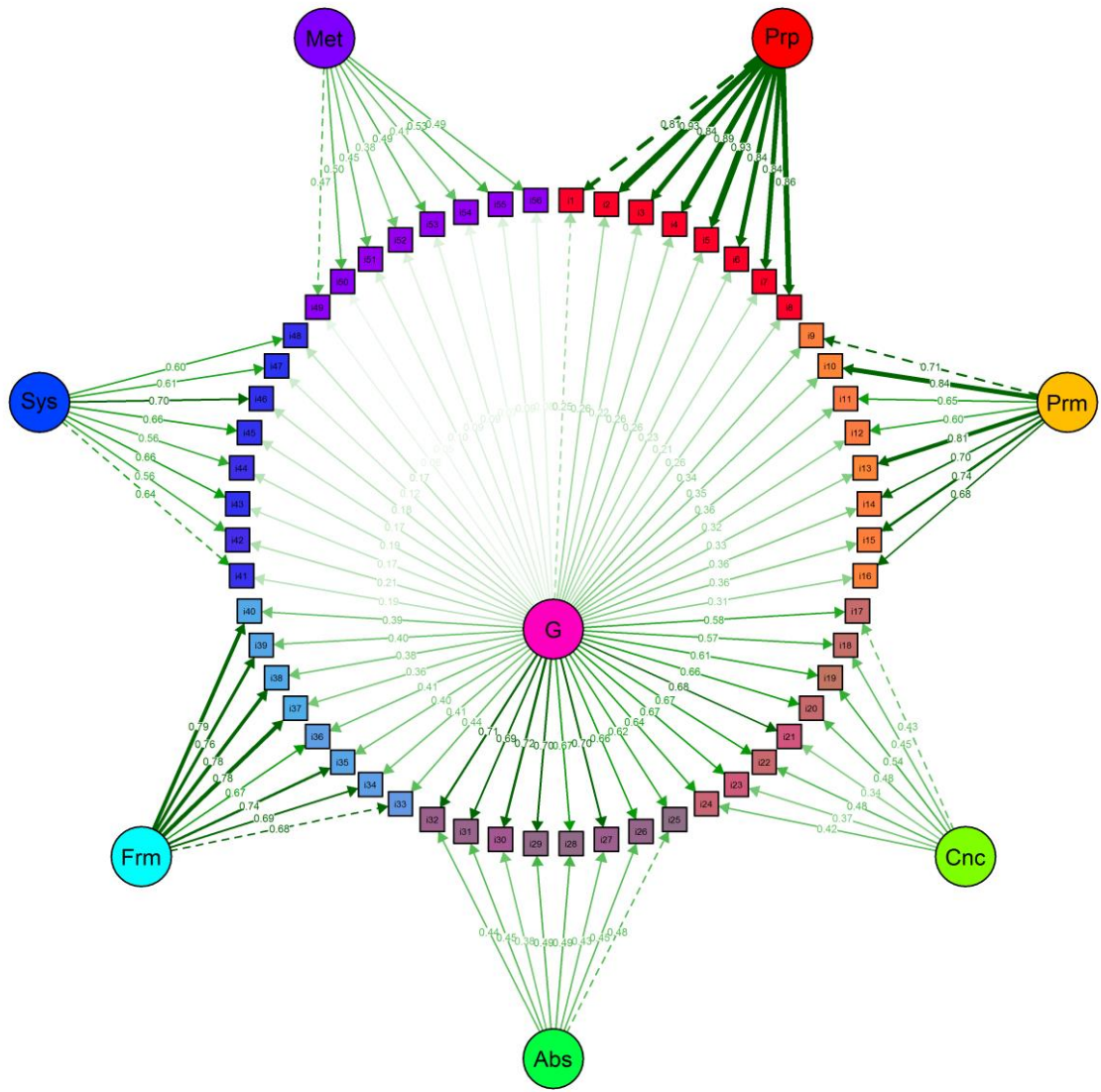

Fig 8. Inductive Reasoning Developmental Test [62] bifactor model. The specific, first order factors correspond to the stages the instrument intended to measure: Prp = Pre-Operational; Prm = Primary; Cnc = Concrete; Abs = Abstract; Frm = Formal; Sys = Systematic; Met = Metasystematic. The general first order factor (G) is the general factor of inductive reasoning.

The EGA was used in the IRDT data and suggested seven dimensions (Fig 9) with its respective items. The nodes represent the items, and the communities, factors or dimensions are colored. It can be seen that the seven dimensions estimated by EGA correspond exactly to the seven first-order factors investigated in the original publication [62]. Parallel analysis, MAP, VSS and BIC and EBIC were used to estimate the number of dimensions in the IRDT data via the *psych* [59] package. Table 3 shows the statistics by number of factors from one to ten, for each method. As can be seen highlighted in bold in Table 3, VSS suggests two factors, Kaiser-Guttman eigenvalue rule suggests six factors, MAP seven, BIC and EBIC ten factors, and parallel analysis four factors. Only MAP suggested the correct number of factors for the IRDT data.

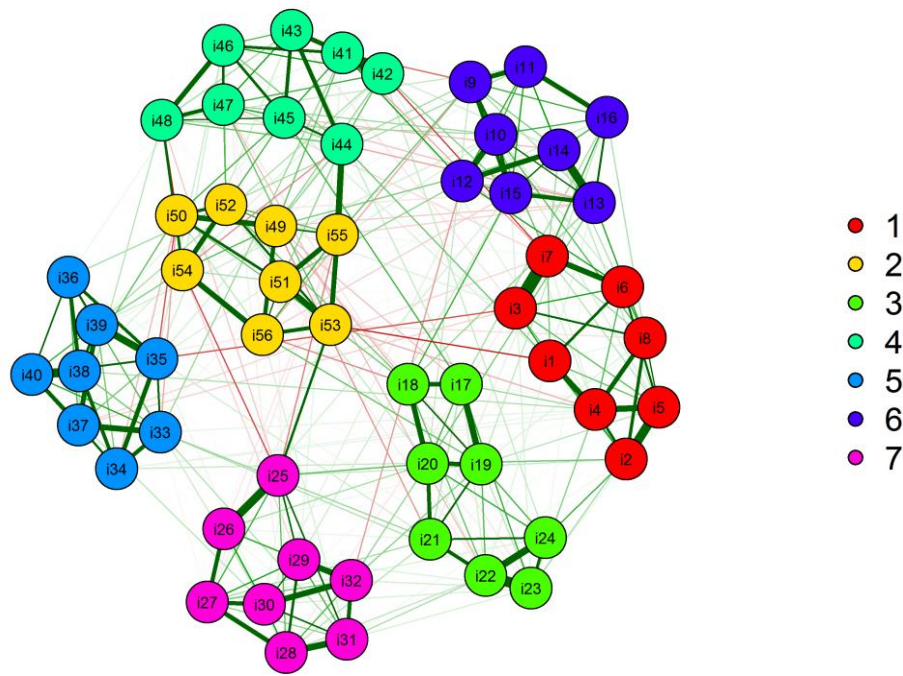

Fig 9. Network of partial correlations estimated during the exploratory graph analysis procedure showing seven latent dimensions in data from the Inductive Reasoning Developmental Test.

**Table 3**
Statistics by each method, from 1 to 10 factors. VSS = Very Simple Structure; BIC = Bayesian Information Criteria; EBIC = Extended Bayesian Information Criteria; MAP = Minimum Average Partial procedure; Kaiser = Kaiser-Guttman eigenvalue rule. The number of factors is chosen as follows: the highest value of the VSS statistic, the lowest value of the MAP, BIC and EBIC statistics, and the last observed eigenvalue greater than the simulated eigenvalue in the parallel analysis.

| Number of factors | VSS | MAP | BIC | EBIC | Kaiser | Parallel Analysis | |
|---|---|---|---|---|---|---|---|
| | | | | | | Observed Eigenvalues | Simulated Eigenvalues |
| 1 | 0.63 | 0.0403 | 36739 | 110227 | 23.17 | 23.17 | 2.69 |
| 2 | **0.68** | 0.0289 | 23717 | 45380 | 7.29 | 7.29 | 2.27 |
| 3 | 0.68 | 0.0192 | 13721 | 19638 | 3.94 | 3.94 | 2.06 |
| 4 | 0.64 | 0.0142 | 8245 | 10762 | 3.02 | **3.02** | **1.80** |
| 5 | 0.66 | 0.0083 | 5326 | -1422 | 1.24 | 1.24 | 1.60 |
| 6 | 0.58 | 0.0078 | 2565 | -4205 | **1.07** | 1.07 | 1.41 |
| 7 | 0.58 | **0.0056** | 1646 | -6297 | 0.85 | 0.85 | 1.22 |
| 8 | 0.59 | 0.006 | 1340 | -6115 | 0.40 | 0.40 | 1.12 |
| 9 | 0.58 | 0.0065 | 1036 | -6011 | 0.33 | 0.33 | 0.98 |
| 10 | 0.58 | 0.0071 | **866** | **-5903** | 0.22 | 0.22 | 0.86 |

**Conclusion**

Estimating the correct number of dimensions in psychological and educational instruments is challenging [1, 2, 3]. We proposed a new method for assessing the number of dimensions in psychological data, which has been derived from the growing field of network psychometrics in which network models are used to model the covariance structure. We term this method *exploratory graph analysis* (EGA), and showed in simulation studies that the method performed comparable to parallel analysis in most cases, and better with multiple strongly correlated latent factors. In addition, EGA automatically identifies which items indicate the retrieved dimensions. We showcased EGA on an empirical dataset of the Inductive Reasoning Developmental Test.

As shown in our simulation study, EGA performed comparable to parallel analysis, EBIC, eBIC and to Kaiser-Guttman rule in a number of situations, especially when the number of factors was two. However, EGA outperformed all methods when the number of items per factor was five and the correlation between factors were high in the four-factor structure. In general, EGA outperformed the other methods in the four factor structure, with a general mean accuracy of 89%, and was the technique whose accuracy was least affected by the conditions investigated in this paper, as shown by the ANOVA's partial eta squared effect size in Table 2. The large differences in the four factors X high correlation X five indicators condition is remarkable, especially compared to the results of the 10 indicator condition. Future simulation studies should confirm if these results can be replicated. Not taking this condition into account, EGA performs comparable to PA over all other conditions with the added benefit of returning which items indicate each dimension.

A surprising evidence appeared in our results: The Kaiser-Guttman eigenvalue greater-than-one rule was better than some researchers would expect [24, 32]. It presented the third best mean accuracy for the two-factor structure (Mean = 86%; SD = 35%) and for the four-factor structure (Mean = 76%, SD = 43%), only losing to parallel analysis (Mean $_{Two-Factors}$ = 97%, SD $_{Two-Factors}$ = 16%; Mean $_{Four-Factors}$ = 80%, SD $_{Four-Factors}$ = 40%) and EGA (Mean $_{Two-Factors}$ = 97%, SD $_{Two-Factors}$ = 19%; Mean $_{Four-Factors}$ = 89%, SD $_{Four-Factors}$ = 31%). However, the Kaiser-Guttman rule suffer from the same issues of parallel analysis, i.e. its accuracy is very low when the correlation between factors is high and the number of items per factor is low. Another results worth pointing refers to the poor performance of VSS, that was the technique less accurate to estimate the number of factors. It should be noticed that while choosing a method to investigate the number of underlying dimensions of a given dataset or instrument, one needs to consider the strengths and weaknesses of each technique, reviewing the scientific literature in order to see the conditions they work the best and the conditions they fail, as well as considering the assumptions of each method. For example, VSS seeks a very simple structure, making very rigid assumptions, that will be met only in a limited number of cases. Both the results of simulation studies and the careful analysis of the underlying assumptions of each method should be considered in order to make a substantiated decision regarding which technique to use.

It is important to note that we have used a very pragmatic approach in our study, since the goal was to investigate whether different procedures can detect the number of simulated dimensions. This is an important part of the development of new quantitative methods aiming to identify the number of dimensions or factors underlying a given instrument or dataset. It is also relevant in order to detect in which conditions the available techniques work the best, in which conditions they should be used carefully and under which circumstances they fail. However, detecting the correct number of factors is only possible for simulated data. Real data allow for several solutions, often similar, especially if one varies the decision criterion. The role of quantitative techniques is to provide support in the quest for understanding the data, supported by careful theoretical analysis, in order to arrive at a solution that is robust both from a quantitative and from a theoretical point of view.

As this is the first study presenting EGA and comparing it to other methods, it has important limitations that should be addressed in future research. Future research should investigate the robustness of EGA to estimate the correct number of dimensions if the data is not multivariate normal, as well as compare it to the well-known and used technique of the Scree-Plot. Also, it would be important to verify the accuracy of other community detection algorithm, besides the walk-trap algorithm currently used in the EGA procedure, in the identification of clusters in undirected weighted networks. A similar investigation was published by Yang, Algesheimer and Tessone [69], which showed the walk-trap algorithm as one of the most accurate ones. However, Yang, Algesheimer and Tessone [69] investigated the accuracy of community detection algorithms for very large undirected weighted networks (with more than 1,000 nodes), which is not the usual number of variables in psychological or educational researches involving the use of tests and/or questionnaires.

There are at least other four things to investigate further. The first two are how EGA works for different levels of factor loadings and for different type of items (polytomous and continuous). It is also important to investigate if the findings of the current paper can be replicated in scenarios involving only one factor. Finally, future research should investigate both the communalities and the proportion of explained variance of the dimensional structure suggested by EGA, especially when using real data. We expect that, in spite of the relevant open questions briefly pointed above, EGA can be used in real datasets. It outperformed other methods, including the very well-known and widely used parallel analysis and minimum average partial procedure, when the number of factors were equal to four, the number of items was five and the correlation between factors were high. In a nutshell, EGA can help with an issue that have been challenging researchers since the beginning of scientific psychological testing. The findings of the current paper may be the solution that Keith, Caemmerer and Reynolds [18] was looking for when they investigated if the available methods underestimates or overestimates the number of factors in intelligence researches. In face of the problems with parallel analysis and MAP, they pointed that a possible solution could be found in formal and informal theory in research with cognitive tests. We can argue that a possible solution is the use of EGA in intelligence like data.